\documentclass[twocolumn]{aastex63}

\usepackage{verbatim}
\usepackage{gensymb}
\usepackage{graphicx}
\shorttitle{Radio Sources in Globular Clusters}
\shortauthors{Shishkovsky et al.}

\begin{document}

\title{The MAVERIC Survey: Radio catalogs and source counts from deep Very Large Array imaging of 25 Galactic globular clusters}

\author[0000-0003-0286-7858]{Laura Shishkovsky}
\affiliation{Center for Data Intensive and Time Domain Astronomy,
Department of Physics and Astronomy,
 Michigan State University,
    East Lansing, MI 48824, USA}

\correspondingauthor{J. Strader}
\email{straderj@msu.edu}
\author[0000-0002-1468-9668]{Jay Strader}
\affiliation{Center for Data Intensive and Time Domain Astronomy,
Department of Physics and Astronomy,
 Michigan State University,
    East Lansing, MI 48824, USA}

\author[0000-0002-8400-3705]{Laura Chomiuk}
\affiliation{Center for Data Intensive and Time Domain Astronomy,
Department of Physics and Astronomy,
 Michigan State University,
    East Lansing, MI 48824, USA}

\author[0000-0002-4039-6703]{Evangelia Tremou}
\affil{LESIA, Observatoire de Paris, CNRS, PSL, SU/UPD, Meudon, France}
   
\author[0000-0003-4553-4607]{Vlad Tudor}
\affiliation{International Centre for Radio Astronomy Research,
 Curtin University,
GPO Box U1987, Perth, WA 6845, Australia}

\author[0000-0003-3124-2814]{James C.A. Miller-Jones}
\affiliation{International Centre for Radio Astronomy Research,
 Curtin University,
GPO Box U1987, Perth, WA 6845, Australia}

\author[0000-0003-2506-6041]{Arash Bahramian}
\affiliation{International Centre for Radio Astronomy Research,
 Curtin University,
GPO Box U1987, Perth, WA 6845, Australia}
\author[0000-0003-3944-6109]{Craig O. Heinke}
\affiliation{Department of Physics, University of Alberta, CCIS 4-181, Edmonton, AB T6G 2E1, Canada}

\author[0000-0003-0976-4755]{Thomas J. Maccarone}
\affiliation{Department of Physics \& Astronomy,
Texas Tech University,
Box 41051, Lubbock, TX 79409-1051, USA}

\author[0000-0001-6682-916X]{Gregory R. Sivakoff}
\affiliation{Department of Physics, University of Alberta, CCIS 4-181, Edmonton, AB T6G 2E1, Canada}

\begin{abstract}
The MAVERIC survey is the first deep radio continuum imaging survey of Milky Way globular clusters, with a central goal of finding and classifying accreting compact binaries, including stellar-mass black holes. Here we present radio source catalogs for 25 clusters with ultra-deep Karl G. Jansky Very Large Array observations.
The median observing time was 10 hr per cluster, resulting in typical rms sensitivities of 2.3 and 2.1 $\mu$Jy per beam at central frequencies of 5.0 and 7.2 GHz, respectively. We detect nearly 1300 sources in our survey at $5\sigma$, and while many of these are likely to be background sources, we also find strong evidence for an excess of radio sources in some clusters. The radio spectral index distribution of sources in the cluster cores differs from the background, and shows a bimodal distribution. We tentatively classify the steep-spectrum sources (those much brighter at 5.0 GHz) as millisecond pulsars and the flat-spectrum sources as compact or other kinds of binaries. These provisional classifications will be solidified with the future addition of X-ray and optical data. The outer regions of our images represent a deep, relatively wide field ($\sim 0.4$
deg$^{2}$) and high resolution C band background survey, and we present source counts calculated for this area. We also release radio continuum images for these 25 clusters to the community.
\end{abstract}

\section{Introduction}
Radio observations of globular clusters (GCs) have been central to studies of compact binaries in GCs for decades. The first radio surveys of GCs were motivated by detection of variable X-ray sources in GCs by \emph{Uhuru} and \emph{OSO-7} \citep{1975clark_mark_li}. There was immediate speculation that these cluster sources were formed dynamically rather than through the evolution of primordial binaries as for field X-ray sources (\citealt{1975clark}; \citealt{1975fab_prin_rees}; \citealt{1975sutantyo}). These radio surveys (e.g., \citealt{johnson1976,johnsonetal1977,roodetal1978,gopalkrishnaetal1980}) were limited to relatively bright sources (typically $> 1$ mJy) and in many cases the sources were found to be more likely associated with the background than with the cluster \citep{birkinshawetal1982}.

Following the discovery of millisecond pulsars, it was speculated that such sources might be abundant in GCs. This led to radio continuum observations with the Very Large Array of a variety of clusters at different angular resolutions, but with few clear detections of candidate pulsars (e.g., \citealt{hamiltonetal1985}). Subsequent work using low-resolution radio imaging showed that a number of clusters had steep spectrum ($-2\lesssim\alpha\lesssim-1$, for flux density {$S_{\nu} \propto \nu^{\alpha}$}) integrated radio flux in their cores, consistent with the idea that many GCs do indeed host large populations of pulsars (\citealt{1990fruch_goss}; \citealt{2000fruch_goss}). Higher-resolution 1.4 GHz imaging of a small sample of well-studied clusters such as M15 identified both known pulsars and low-mass X-ray binaries, as well as sources with no known counterparts \citep{kulkarnietal1990,1991john_kulk_goss}. Subsequent to these studies, the successful discovery and timing of many GC pulsars with Parkes, Arecibo, and later the Green Bank Telescope (e.g., \citealt{robinson1995}; \citealt{hessels2007}; \citealt{lynch2011}) bore out the promise of the radio imaging results.

While pulsars are expected to be abundant in deep low-frequency radio continuum imaging of GCs (e.g., \citealt{camilo_rasio05}; \citealt{ransom2008}), other classes of binaries readily observed at other wavelengths (such as in the X-ray) should also be detectable in the radio. Low-mass X-ray binaries (LMXBs), in which a neutron star or black hole accretes from a Roche lobe-filling companion, are abundant within GCs. In field neutron star LMXBs, radio emission associated with non-thermal jets has been observed at X-ray luminosities of $L_X \gtrsim 10^{34}$ erg s$^{-1}$ (e.g., \citealt{tudor_etal2017}), and down to even lower luminosities of $L_X \sim 10^{33}$ erg s$^{-1}$ for a few ``transitional" millisecond pulsars (e.g., \citealt{delleretal2015}), though the origin of the radio emission in this state is less clear \citep{bogdanovetal2018}. Black hole LMXBs have been observed in the radio even down to quiescent luminosities of $L_X \sim 10^{30}$ erg s$^{-1}$ (e.g., \citealt{galloetal2014}), with the emission thought to come from partially self-absorbed synchrotron radiation from a compact jet (\citealt{1979blan_konigl}, \citealt{hjellming_johnston1988}). Other binaries with lower typical X-ray luminosities, such as cataclysmic variables, active binaries, and even exotic systems such as ``white dwarf pulsars" are also detectable with sufficiently sensitive imaging (e.g., \citealt{drake1992,abadasimonetal1993,coppejans2015,marsh2016}).

Given the depth of modern X-ray images (from \emph{Chandra} or \emph{XMM}) it is worthwhile to consider the utility of new radio imaging observations of GCs. One motivation is the search for radio emission from putative accreting intermediate-mass black holes (e.g., \citealt{maccarone2004,tremouetal2018}). Another is the possibility of distinguishing between neutron star and black hole primaries in LMXBs: on average, accreting black holes show more luminous radio emission than neutron stars at a fixed $L_X$ (e.g., \citealt{galloetal2014,galloetal2018}), even though the X-ray spectra look very similar. Nearly all X-ray bright LMXBs in GCs are confirmed to host neutron stars, but for many of the fainter sources, the identity of the accretor is unknown, except in those cases where thermal emission from the surface of the neutron star is detectable or multi-wavelength information is available (e.g., \citealt{heinkeetal2003}).

For many years it was thought unlikely that GCs hosted stellar-mass black holes in any substantial quantity, with their ejection the natural outcome of mass segregation and violent interactions in the cluster core \citep{sigurdssonetal1993,kulkarnietal1993}. This view was revisited starting about a decade ago with the discovery of a compelling stellar-mass black hole candidate in an extragalactic GC (e.g., \citealt{2007maccarone_etal}; \citealt{2008zepf_etal}). 

The subsequent upgrade of the Karl G.~Jansky Very Large Array (VLA), yielding a great increase in its radio continuum sensitivity, enabled the possibility of a wide-scale search for radio emission from quiescent black hole LMXBs in GCs. Our pilot surveys with the VLA and the Australia Telescope Compact Array (ATCA) successfully uncovered stellar-mass black hole candidates in the GCs M22 \citep{straderetal2012}, M62 \citep{chomiuketal2013}, and 47 Tuc \citep{millerjonesetal2015}. We then initiated MAVERIC (Milky Way ATCA VLA Exploration of Radio Sources in Clusters) as a systematic survey for radio sources in 50 GCs with VLA and ATCA. The broader scientific implications of this project have accelerated in the last few years with the discovery that close black hole--black hole binaries are common in the local universe \citep{abbottetal2016}, with dynamical formation in GCs a possibility to account for some or many of these systems \citep{rodriguezetal2016,chatterjeeetal2017}. There has been a similar salvo of theoretical work on the presence of stellar-mass black holes in GCs, now arguing that some clusters could host tens to hundreds of black holes at the present day (e.g., \citealt{Mackey08,Sippel13,morscheretal2015,kremeretal2018,weatherfordetal2018})

While the ultimate goal of the survey is to identify, follow-up, and classify a wide variety of binaries, including perhaps accreting stellar-mass black holes, it will not be possible in every case to definitively assess a source's classification. Even if a given source is associated with a compact object, a binary could have a sufficiently faint donor star that dynamical confirmation of the nature of the accretor is beyond the capability of present-day instrumentation. Therefore, in addition to these detailed investigations, it is useful to take a \emph{global} view of the radio source populations in our radio images. 

Here, as an intermediate product of our overall scientific goal to assess the population of radio sources, especially black hole candidates, in Galactic GCs, we present a radio source catalog compiled from our deep VLA observations of 25 Milky Way globular clusters (the GCs with ATCA observations are presented in a companion paper by Tudor et al., in preparation). We use these catalogs to do a source count and spectral index analysis of the radio continuum sources in our images, and show evidence for a significant population of individual radio sources in both our combined sample and in some individual GCs.

This paper is arranged as follows. In Section 2 we describe our  data reduction procedures. Section 3 outlines our source finding method and spectral index analysis. Section 4 contains our main source count and spectral index results. We summarize and conclude in Section 5.

\section{Radio Observations \& Data Reduction}

\subsection{Sample and Observations}

\begin{figure}[t!]
\includegraphics[width=0.47\textwidth]{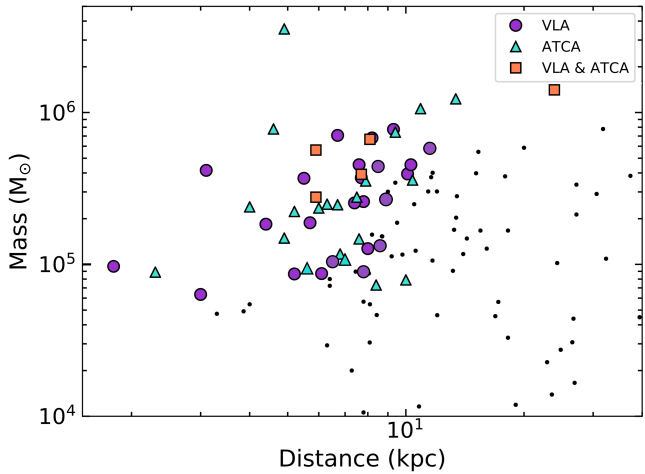}
\caption{{GC mass vs.~distance for Galactic GCs, showing the selection of the MAVERIC (VLA and ATCA) GC sample with {masses $\gtrsim 10^5 M_{\odot}$ and distance $\lesssim 9$ kpc}, with an extension to massive distant GCs for intermediate mass black hole searches. The VLA sample is shown by the purple circles and the ATCA sample (Tudor et al., in preparation) by the turquoise triangles. Orange squares are the five clusters observed by both the VLA and ATCA. Black points are other Galactic GCs \citep{2018baumgardt}. For the VLA sample, the cluster distances are listed in Table \ref{tab:vla_im} (see \citealt{tremouetal2018} for a full set of references) and the masses are taken from \citet{2018baumgardt}.}
\label{fig:mass_dist}}
\end{figure}

The initial MAVERIC cluster sample was selected to include all Galactic GCs with masses $>$ 10${^5}$ M$_{\odot}$ and distances $< 9$ kpc (Figure \ref{fig:mass_dist}). These limits were designed to include massive clusters more likely to host a black hole population, and clusters close enough that we had sufficient sensitivity to detect radio emission akin to the candidate quiescent stellar-mass black holes in M22 \citep{straderetal2012}. We also added a few more massive GCs at larger distances, since these GCs are potential hosts for intermediate-mass black holes. More recent mass and distance determinations have moved a subset of the sample beyond these nominal mass and distance limits (see Figure 1), but the spirit of the selection is intact.

Owing to its higher sensitivity, we observed as much of the sample as possible with the VLA, reserving ATCA for the more southerly clusters not accessible to the VLA. There are 50 clusters in the final sample. The 25 clusters for which reasonably homogeneous VLA datasets have been completed are presented in this paper; we expect to add about 7 additional objects in future papers. 

Pilot observations for the survey were made as part of National Radio Astronomy Observatory (NRAO) programs VLA/10C-109 (P.I.~Chomiuk) in 2011 and  VLA/12B-073 (P.I.~Strader) in 2012. The main survey was approved as an NRAO Large Program (project codes VLA/13B-014 and VLA/15A-100), with observations made in 2014 and 2015, respectively. Table \ref{tab:vla_epoch} lists the observation blocks for each cluster. The observations were primarily made in the most-extended A configuration, though a subset of southern clusters were observed in BnA or BnA to A ``move time". The goal and median observing time per cluster was 10 hr, though a few clusters had a total observation time shorter or longer than this goal. Of this 10 hr of total observing time, the median time on source was 7.4 hr. Most of the off source time was allocated to the observation of phase calibrators, with a median cycle time of about 10 min. The observing efficiency was determined primarily by the length of the observed blocks. Table \ref{tab:vla_im} lists the total and on source time for each cluster.

We made a few observations that are not used in this paper: these were all 
taken in move time between A and D configurations. In the course of verifying the catalogs in this paper, we found that the flux densities of apparent point sources were lower in these data compared to non-move time VLA data or ATCA data for the same clusters. We tried many experiments
to account for these discrepancies and were unable to do so, other than to say they are consistent with substantial decorrelation (up to a factor of 2) in the A to D move time data. These data are excluded from this paper. The main effect of this is to remove Liller 1 and NGC 6522 from our present sample.\footnote{We also note that the VLA intermediate-mass black hole upper limits from \citet{tremouetal2018} for only these two clusters should be increased by about 30\%. While this has no effect on the overall conclusions of that paper---and indeed ATCA data for these clusters show no evidence of a central source that could be an intermediate-mass black hole---these clusters should be revisited when higher-quality data are available.}

For all clusters we observed with the C band receiver, including full polarization products. Data taken in 2014 or earlier used an 8-bit setup with two 1-GHz basebands centered at 5.0 and 7.4 GHz, while the 2015 data used 3-bit receivers with two 2-GHz basebands centered at 5.0 and 7.0 GHz. However, the final central wavelengths depend on the details of the flagging of radio frequency interference, which varies among the sample (see Table \ref{tab:vla_im}). For simplicity, for the remainder of the paper we refer to the lower baseband as 5.0 GHz and the upper baseband as 7.2 GHz, even if the central frequencies differ slightly from these values.

\subsection{Data Reduction}

The data for both cluster frequency bands were separately flagged and calibrated according to standard procedures with either \textit{Common Astronomy Software Application} (CASA; \citealt{2007mcmullin_casa}) version 4.2.2, or \textit{Astronomical Image Processing System} (AIPS; \citealt{2003greisen_aips}). Before imaging, the reduced visibilities from each observational epoch were stacked to maximize the sensitivity in each baseband. The combined bands were then imaged separately, with minimum fields of view (diameters) of 11{\arcmin} at 5.0 GHz and 7.5{\arcmin} at 7.2 GHz. These were chosen to match the full width at half-maximum (FWHM) of the primary beam at the lowest frequency of each baseband (4.0 or 6.0 GHz, respectively). Pixel sizes were set to $0.08''$ at 5.0 GHz and $0.06''$ at 7.2 GHz to adequately sample the synthesized beams of each frequency range. In AIPS, the data were imaged using IMAGR, Briggs weighting with a robust parameter of 1, frequency-independent deconvolution, and facets to account for wide-field effects. With data imaged in CASA/clean, nterms=2 was used to account for non-zero spectral indices of sources in the field. A primary beam response correction was applied to each subband image assuming the average frequency in that subband. The synthesized beam size for each image, as well as the image rms flux density values in $\mu$Jy per beam, are listed in Table \ref{tab:vla_im}. The median rms values for the low and high frequency basebands across all clusters were 2.3 and 2.1 $\mu$Jy per beam, respectively. We used self-calibration on the few clusters for which there was sufficient flux ($\gtrsim 5$ mJy), typically with a single pass of phase self-calibration with a solution interval of 10 min.

The final reduced images for each cluster are hosted as science-ready data products at NRAO.

\section{Source Finding Procedure \& Analysis}

We do source finding on the individual baseband images
using \textsc{Aegean} (\citealt{2012hancock_aeg}; \citealt{2018hancock_aeg}) and the associated program \textsc{BANE}.
This latter routine creates local background and rms images which are used as input to \textsc{Aegean}. \textsc{Aegean} searches for individual pixels above a threshold to seed a potential source, then grows this source using pixels above a lower threshold. With a goal of having a final catalog of sources of 5$\sigma$ significance, after experimentation, we settled on a seed threshold of $4.5\sigma$ and a grow (``flood") threshold of $3\sigma$. The final significance cut was made after final flux density fitting (see below).

Given the modest extent of most of the GCs (median half-light radius 1.3\arcmin) and the desire to have spectral information for source classification, in both basebands we restrict our source finding to a radius of $3.7\arcmin$ around the cluster center. This value is the typical half-width at half maximum of the upper baseband, and is sufficiently large to cover the area within the half-light radii of all but one of the sample GCs (M4: $r_h$ = 4.33\arcmin).

All detections were visually inspected. The main changes made after this inspection were to remove a few instances of obvious artifacts and to note any extended sources that had been split into multiple detections. These were replaced by a single source positioned at the most central or compact location of the emission. 

Given the high resolution of our images and the expectation that all cluster sources of interest will be unresolved, we performed the final source fitting in a different manner. The detection catalogs were fed as input to the AIPS task {\tt JMFIT}, which fits a model of the beam to a small 20 pixel box (1.6\arcsec\ for lower baseband; 1.2\arcsec\ for upper baseband) around each detection. This fitting gave a flux density and uncertainty and position and uncertainty for each source. Sources were removed if $< 5\sigma$ after this process. After experimentation, we also removed any sources whose position changed by more than 4 pixels from the initial detection. We then matched the baseband source lists. The final catalog for each GC contains all sources detected at $5\sigma$ at either frequency.

The formal astrometric uncertainties in source positions are related to the beam FWHM and the signal-to-noise (SNR) of the detection as $\sim$ FWHM/(2 SNR), where the approximation is due to the typically correlated nature of noise in radio images, among other factors \citep{condon1997}. However, calibration uncertainties typically limit the astrometric precision of the VLA to $\sim 10\%$ of the beam FWHM.\footnote{https://science.nrao.edu/facilities/vla/docs/manuals/oss/ performance/positional-accuracy} In the absence of additional external tests of the astrometric precision, espescially at the level of $\lesssim 0.1\arcsec$, we adopt this value as a floor. Future tests of the quality of our absolute astrometry would be valuable.

\subsection{Spectral Index Analysis}

The radio spectral indices contain important information about the nature of the sources, and some information is available even for those sources without $5\sigma$ detections at both frequencies. For such sources, we force fit the location of the significant source at the other frequency using {\tt JMFIT}. If the resulting fit yields a $3\sigma$ detection at a location within 4 pixels of the original location, this is retained as a valid detection of the same source at the other frequency. If these conditions are not met, the $3\sigma$ upper limit is given instead. For these cases with limits, we also report the exact flux density at the corresponding pixel (in case a reader is interested), but we only use the upper limits for the subsequent analysis in this paper.

We modeled spectral indices assuming a power law with the flux density {$S_{\nu} \propto \nu^{\alpha}$.} We used a Bayesian Markov Chain Monte Carlo code to fit these values, which self-consistently includes both the uncertainties on the flux densities and the $3\sigma$ upper limits, if present. We assumed a flat prior over the range $\alpha = -3.5$ to +1.5. This prior has little effect for brighter sources with measurements at both flux densities but a larger effect for very faint sources, or those with a single detection. In the individual source tables, if there are detections in both bands, we report the median of the posterior as the best point estimate of the spectral index, equivalent to $1\sigma$ limits (containing 68.3\% of the posterior density around the median). If there is only a detection in a single band, we report a $3\sigma$ lower or upper limit on the spectral index.

\subsection{Final Catalogs}

\begin{figure*}[!th]
\gridline{\fig{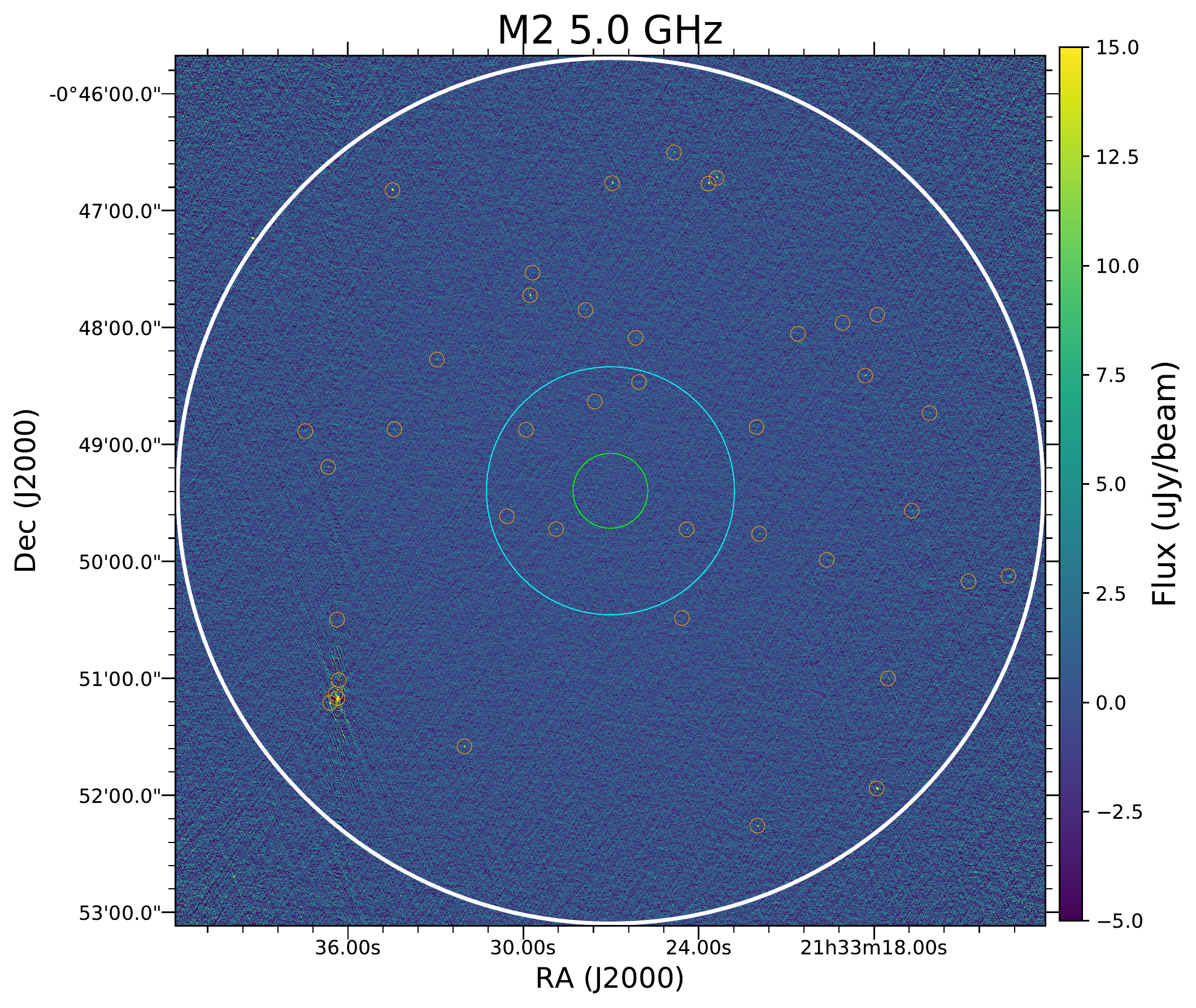}{0.63\textwidth}{\large{(a)}}}
\gridline{\fig{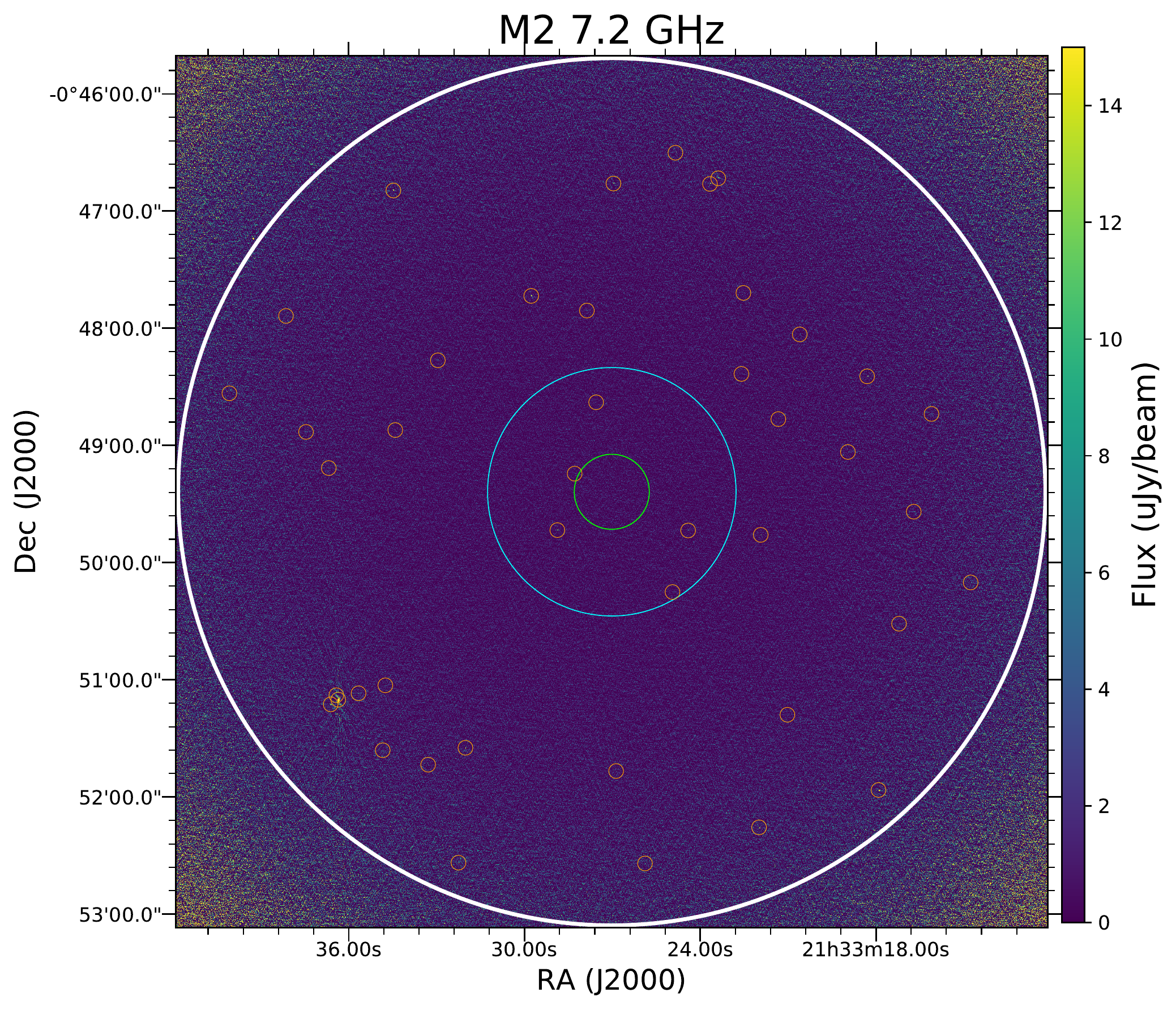}{0.63\textwidth}{\large{(b)}}}
\caption{Final images of M2 at 5.0 GHz (a) and 7.2 GHz (b). The white circle represents the area within our 3.7\arcmin\ total search radius, and the green and blue circles show the areas enclosed by the cluster core radius and half-light radius of M2. The detected sources at each frequency are indicated in the corresponding image by the orange circles.}\label{fig:M2_ims}
\end{figure*}

Our final catalogs are available in machine-readable format. Table \ref{tab:vla_src} shows a sample of the catalog for M2, and the final 5.0 GHz and 7.2 GHz images of M2 are shown in Figure \ref{fig:M2_ims} with the source detection regions overlaid.

A small fraction of the sources in this catalog are known previously, such as bright millisecond pulsars or luminous X-ray binaries. A comprehensive cross-matching with other catalogs and existing data at other wavelengths is a substantial undertaking that we defer to a future paper.

\subsection{Potential Sources of Bias}

There are several sources of bias that can affect the interpretation of our catalogs.

First, the faintest sources in our catalog could potentially be affected by ``flux boosting": sources pushed above their true flux densities by rare positive noise spikes at the source position. We discuss the effects of flux boosting on our source count analysis in \S 4, but in general, flux boosting is unlikely to significantly affect 5$\sigma$ catalogs like the ones in this paper.

A related, but separate issue is that of entirely spurious sources. In the highest resolution images at 7.2 GHz there are of order $2 \times 10^6$ beams within the central $3.7\arcmin$, which would imply about 0.5 false detections per cluster if the noise were perfectly Gaussian (the false detection rate at 5.0 GHz is about half this size). Since real noise is correlated it is reasonable to expect a somewhat higher false detection rate, and indeed in some of the clusters there are $5\sigma$ sources detected at 7.2 GHz without even a $3\sigma$ detection at 5.0 GHz. Sources with such inverted spectra ($\alpha \sim 1.4$) are expected to be rare and hence many or most of these 7.2 GHz-only sources may indeed be spurious. {Frequency-dependent variability, due for example to refractive scintillation (e.g., \citealt{Hancock2019}), could also be relevant for a subset of these sources.} All $5\sigma$ sources are retained in the catalogs but they should be interpreted with these caveats. We expect few spurious sources within the smaller cluster cores, especially at 5.0 GHz.

A separate source of bias is ``resolution bias": the loss of diffuse flux from sources more extended than the synthesized beam. This can result in non-detection of sources whose total flux would place them above the nominal flux density detection limit, or a bias to a lower flux even if detected. While we expect most of the cluster sources to be unresolved, this is not the case for field or background sources. Previous work on radio source counts have estimated the effects of resolution bias, which has the strongest impact on sources below 100 $\mu$Jy, but with a substantial uncertainty both in the size of the effect and how it depends on frequency \citep{windhorstetal1990,windhorstetal1993,huynhetal2020}. This bias may affect the number counts around these flux densities by up to $\sim 25$\%, depending on the assumptions made about the source size distribution \citep{windhorstetal1984,windhorstetal1993,fomalontetal1991}. We also note that we did not attempt to measure the diffuse flux from a small number of extended bright (mostly $>> 1$ mJy) sources. The listed flux densities for these sources solely reflect a point source fit to the core. All sources that appeared to be extended in our imaging, whether bright or not, are marked as such in the catalog.

\section{Source Counts}

A number of radio continuum sources discovered in our survey have already been confirmed as cluster members through optical spectroscopy (e.g., 
\citealt{shishkovskyetal2018}), with compelling multi-wavelength evidence for others (e.g., \citealt{chomiuketal2013}). Such work will eventually be extended to our entire survey in a systematic manner. In the meantime, we consider the radio source counts in the clusters by themselves.

Historically, studies of radio source counts were undertaken with the goal of understanding the evolution of the universe, as their distribution was among the earliest evidence against the existence of uniform cosmic evolution \citep{ryleetal1955,scheuer1957,ryleetal1962}. The suggested sources that dominate the differential counts are radio-loud AGN, galaxies with active star formation, and radio-quiet AGN (e.g., \citealt{condon1984}), with each class contributing more or less as a function of flux density. Bright sources are mostly AGN, but below $\sim 1$ mJy extragalactic source counts exhibit some flattening in the the slope of the Euclidian-normalized differential counts. The origin of this flattening is uncertain, and may be due to radio-quiet AGN, star forming galaxies, or a combination thereof \citep{padovanietal2007,huynhetal2008,smolcicetal2008}.

{Known compact Galactic radio sources at 1.4 GHz (the most common survey frequency)} are primarily composed of \ion{H}{2} regions and nominally confined {near the Galactic plane,} with latitudes $|b| \; {<} \; 5\degree$ (e.g., \citealt{beckeretal1990,griffithetal1993,giveonetal2005}). {In a 5.0 GHz VLA survey of Galactic plane radio sources,} \citet{purcelletal2013} found that bright resolved \ion{H}{2} regions overwhelmingly accounted for the excess of radio sources around $|b| \; = \; 0\degree$, yet only comprised $\sim 25\%$ of their 7$\sigma$ sample (see also \citealt{hoareetal2012}). The \ion{H}{2} regions were distributed very close to the Galactic plane with a scale height of 0.47$\degree$. Their unresolved sources dominated the fainter population, and were distributed evenly over their Galactic latitude coverage ($|b| \; {<} \; 1\degree$), suggesting that the faint compact sources in their sample were predominantly background sources. However, the flux density limits of the \citet{purcelletal2013} study were a factor of $\sim 100$ brighter than the VLA GCs in MAVERIC, so it is quite possible our data are probing a new, fainter regime where the density of Galactic radio continuum sources is yet unknown.

We can make a separate argument that {Galactic plane} sources are unlikely to substantially affect our study: none of our sample GCs are within 1.5$\degree$ of the Galactic plane, and only four (NGC 6440, NGC 6544, NGC 6712, NGC 6760) are within 5$\degree$ of the plane. These four GCs have an average of 18 sources detected at $5\sigma$ at 5.0 GHz in the radial range 2.5--3.7\arcmin\ that we take as the ``background" region. The other VLA GCs with imaging of comparable sensitivity have an average of 18.6 sources per GC in this region. Hence, from our data themselves there is no evidence of a population of sources that is more common at low Galactic latitude, as would be expected for essentially any Galactic population except for one within tens of pc of the Sun. The serendipitous discovery of a 0.2--0.3 mJy Galactic radio continuum source in the foreground (distance $\sim 2.2$ kpc) of the GC M15 ($b = -27$) using very long baseline interferometry \citep{kirstenetal2014,tetarenkoetal2016} shows that such sources do exist at some surface density, but evidently not a level sufficient to measurably affect our results.

The main goal of the source counts in this paper is to allow an estimate, on a cluster-by-cluster basis, of the evidence for a population of radio sources associated with the GC itself. However, we also check our source counts against previous work and models of the faint extragalactic population.

\subsection{Calculation of Radio Source Counts}

We present our radio source counts in the typical differential form normalized to a static Euclidean universe, given by:
${S_{\nu}}^{2.5} { { \frac {dN}{dS}}} ({\text{Jy}}^{1.5}\;{\Omega^{-1}})$. 
Since each image has a different depth, and within each image the sensitivity falls off with radius, the catalogs cannot be directly used to calculate source counts.
Instead, for each cluster, we sort the sources by flux density and radius, and use the aforementioned background and rms images from {\tt BANE} to create sensitivity maps. We use these to calculate the area of the image in which a source could have been detected at $5\sigma$.
These corrected counts, added together for all clusters, make up the full differential source counts for the survey. The total survey area is $\sim 0.4$ deg$^2$, with a median $5\sigma$ depth of $\sim 11\mu$Jy per beam, making it among the deepest C band surveys with this angular coverage.

\begin{figure}[!th]
\gridline{\fig{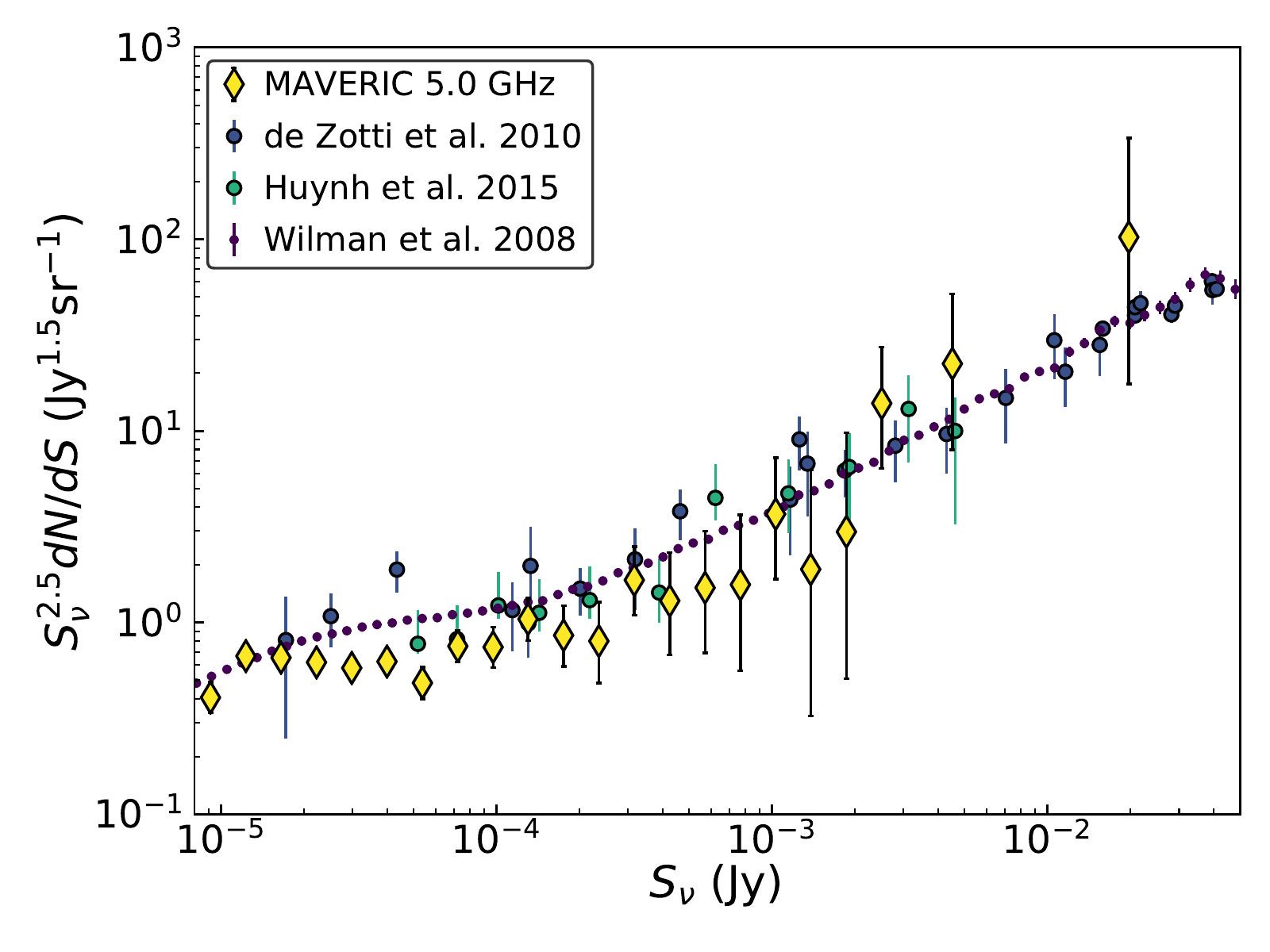}{0.5\textwidth}{\vspace{-0.6cm}\small{(a) 5.0 GHz source counts}}\label{fig:sc5}}
\vspace{-0.5cm}
\gridline{\fig{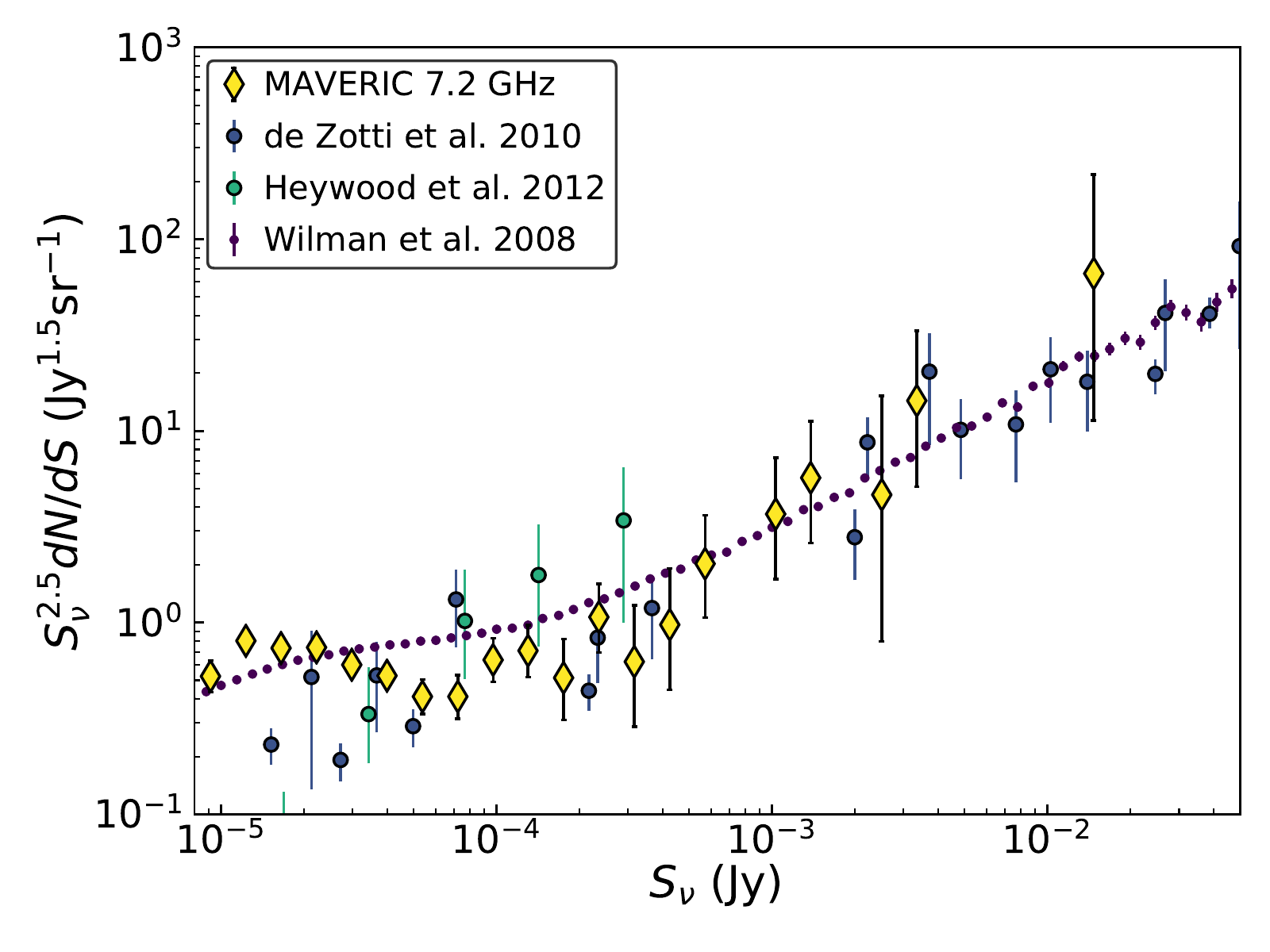}{0.5\textwidth}{\vspace{-0.6cm}\small{(b) 7.2 GHz source counts}}\label{fig:sc7}}
\vspace{-0.3cm}

\caption{Differential radio source counts for our sample at 5.0 GHz (a) and 7.2 GHz (b), normalized for the standard $S^{2.5}$ Euclidian term (yellow diamonds). The other points represent different literature measurements, as described in the text. Our measurements are in general agreement with published values but slightly lower in the flux density range $\sim 50-200$ $\mu$Jy, likely due to us resolving out diffuse flux at our higher resolution.}

\label{fig:scounts}
\end{figure}

These radio source counts are shown in Table \ref{tab:scounts} and plotted in Figure \ref{fig:scounts}. As expected, we observe a modest flattening in the Euclidian-normalized differential source counts below 1 mJy, and perhaps an upturn in the faintest flux bins.
Fainter sources dominate the total counts: 1254 of the total 1267 5$\sigma$ sources are below 700$\mu$Jy.

For comparison, we also plot the source counts from several other deep surveys. At the lower frequency, \citet{huynhetal2015} present source counts from a deep ATCA survey, reaching $43 \mu $Jy ($5\sigma$) over $\sim 0.34$ deg$^{2}$ in the $\it{Chandra}$ Deep Field South. We scale these 5.5 GHz counts to 5.0 GHz using the median spectral index of their sample ($\alpha=-0.58$). \citet{heywoodetal2013} describe their radio catalog derived from deep VLA observations of the William Herschel Deep Field, reaching a sensitivity of 2.5$\mu$Jy. Their 8.4 GHz source counts are included  with the upper band counts, and were scaled to 7.2 GHz using the typical extragalactic spectral index $\alpha$ = --0.7. We also include the older literature compilation by \citet{dezottietal2010}, scaled from 4.8 GHz to 5.0 GHz and 8.4 GHz to 7.2 GHz assuming $\alpha$ = --0.7. Finally, we include the simulations of \citet{wilmanetal2008}, which were derived semi-empirically using existing radio luminosity functions extrapolated down to nanoJy levels. These values, presented at 4.86 GHz, are scaled to both 5.0 GHz and 7.2 GHz, again using $\alpha$ = --0.7.

Overall, given the uncertainties in our new measurements and the scatter among previous source count estimates, our differential source counts are generally consistent with previous work, at least down to $\sim 100\mu$Jy. We note that somewhat lower counts for brighter sources might be expected from our survey, given our high resolution and limited treatment of extended emission.

Fainter than $100\mu$Jy, our source counts have much smaller Poissonian uncertainties than the comparison studies. The differential source counts are flat down to the $\sim 11\mu$Jy $5\sigma$ survey limit at 5.0 GHz and show a small upturn at 7.2 GHz. Since most sources are not expected to have inverted radio spectra, we attribute this 7.2 GHz upturn primarily to spurious sources rather than real sources, but this point should be revisited in the context of future follow-up and other deep surveys.

\subsection{Cluster Radio Source Excess}

To determine whether individual GCs show an excess of radio sources, we choose a somewhat conservative approach with a minimum set of assumptions. Since most sources have higher flux densities at 5.0 GHz than 7.2 GHz, we only carry out the following analysis at 5.0 GHz.

Because of cosmic variance, it is better to use a local estimate of the background, rather than the global source counts discussed above. After examination of the radial distribution of sources in each cluster, we found that radii beyond 2.5\arcmin\ are strongly dominated by non-cluster sources in most or all GCs. Hence we chose an annulus of 2.5--3.7\arcmin\ as a clean, relatively large area in which a local background can be estimated. For each cluster, we estimated the areal density of background sources using a sensitivity analysis similar to that as for the source counts. 

We took two bracketing approaches for defining a likely set of radio sources that could be associated with the GC: the sources within the core radius or within the half-light radius. The former is guided by the observed central mass segregation of X-ray populations within GCs (e.g., \citealt{grindlay1984}; \citealt{verbunt1987}; \citealt{grindlay2002}; \citealt{verbunt2006}). Cluster X-ray sources are mostly binary stars or their progeny: the relatively massive (and hence centrally concentrated) objects one would expect to be observable as radio continuum sources. If the radio sources instead follow the cluster stars---an unlikely limiting case since nearly all single cluster stars are undetectable in the radio with current facilities (e.g., \citealt{Maccaroneetal2012})---then half would be expected to be contained within the half-light radius. 

To determine the ``excess" number of sources within the core or half-light radius of each GC, we first calculated the expected number of background sources by scaling the sensitivity-corrected background density to the comparison GC area. We then subtracted this from the sensitivity-corrected observed number of sources in the core or half-light radius. This latter correction had little effect on the results, with two exceptions discussed below. Uncertainties in all quantities were determined by bootstrapping.

For the few GCs with half-light radii larger than 2.5\arcmin\ (M4, M22, and M55), we use 2.5\arcmin\ as the boundary between the background and the inner area we take to be associated with the GC. While we cannot be entirely sure that there are zero sources associated with the GC outside 2.5\arcmin\ in these GCs, they have an average background source density comparable to that of the other clusters, suggesting this assumption is reasonable for our first-order analysis. In our final catalog, there are 54 sources detected at 5.0 GHz within the core radius of a cluster, 201 within the half-light radius (this includes the 54 sources with the core radius), and 435 in the 2.5--3.7\arcmin\ background region.

\begin{figure*}[!th]
\gridline{\fig{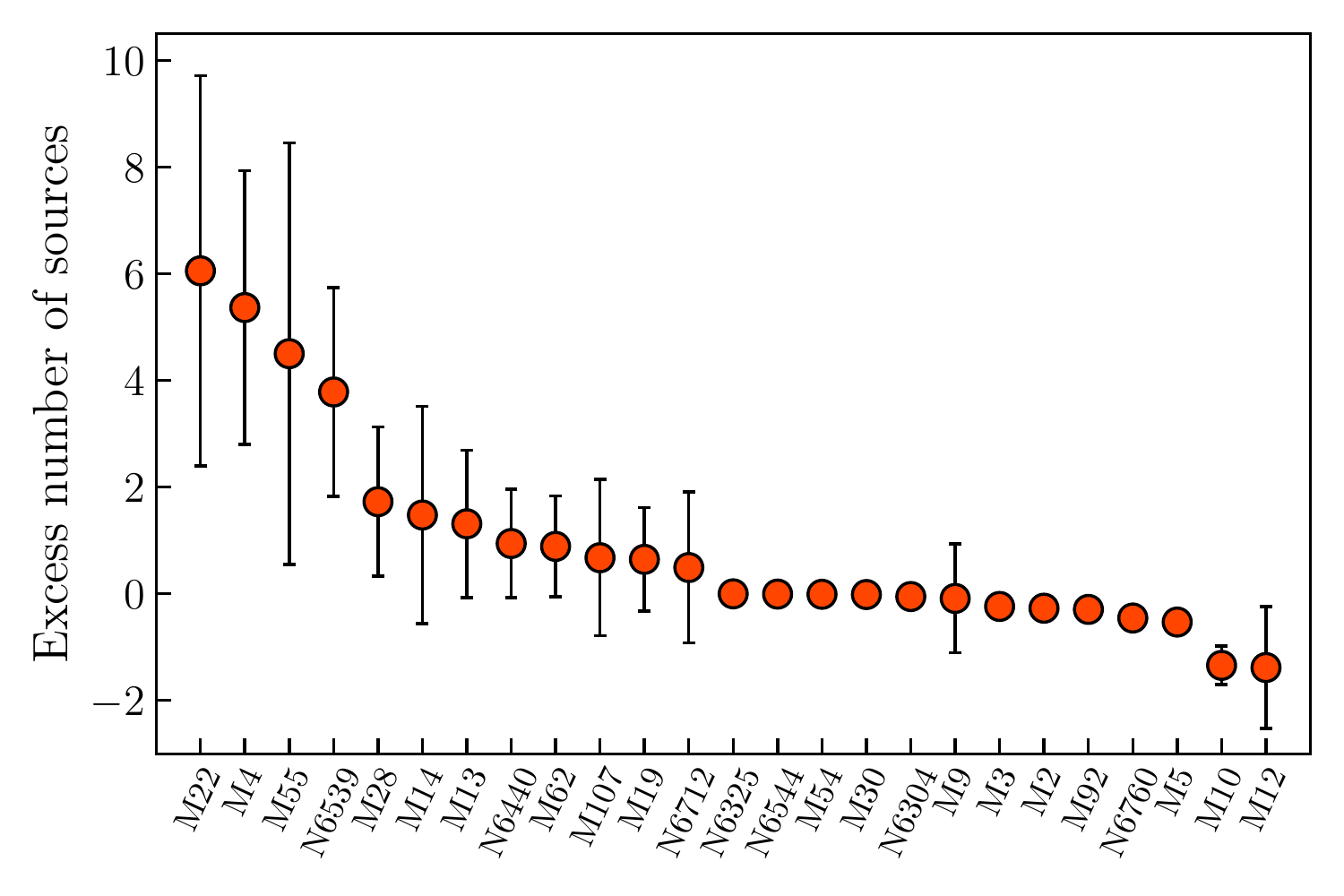}{0.7\textwidth}{\large{(a) Core radius excess}}}
\gridline{\fig{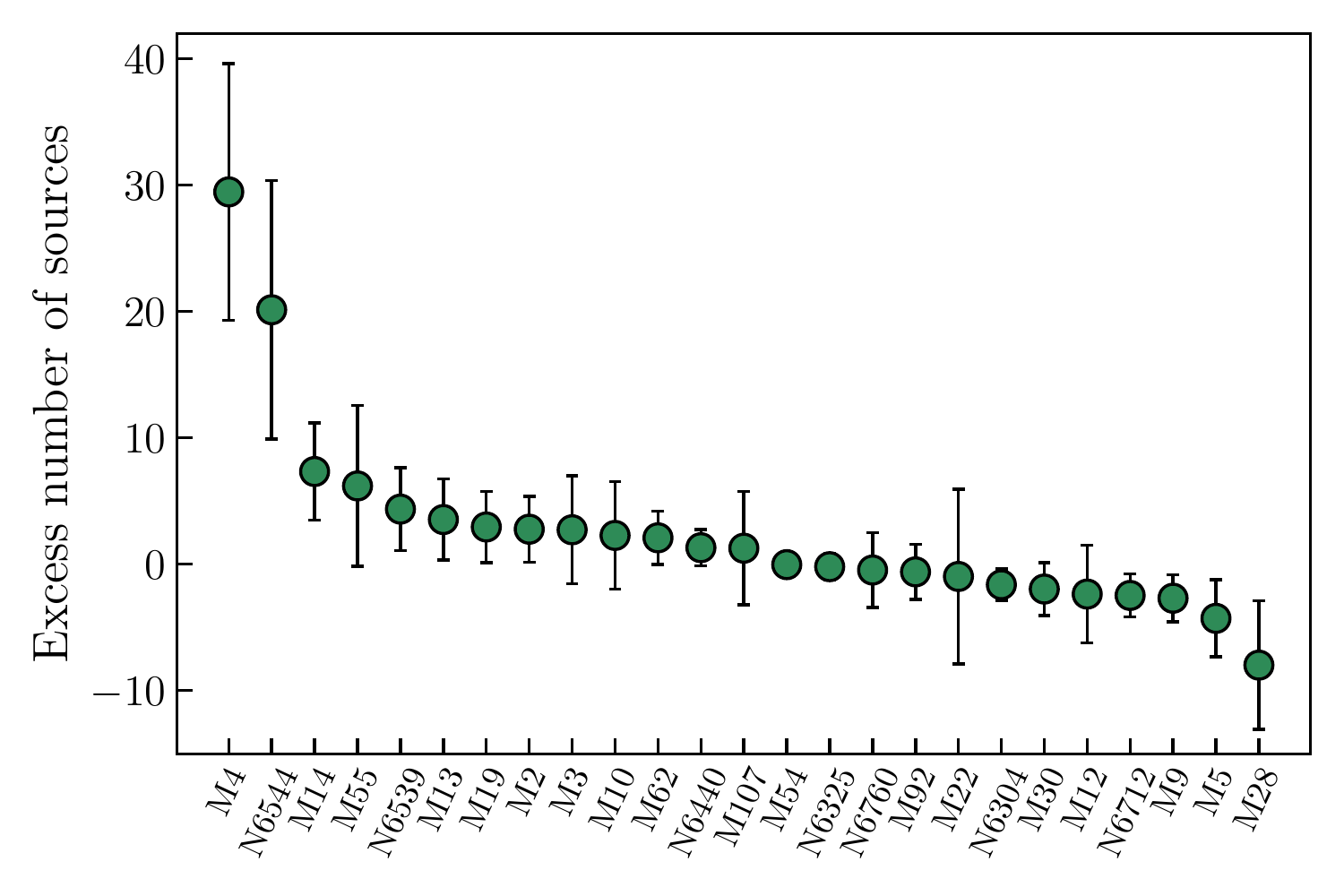}{0.7\textwidth}{\large{(b) Half-light radius excess}}}
\caption{``Excess" of 5.0 GHz radio sources 
within the core radius (top panel) or half-light radius (bottom panel) of each GC in our sample, compared to the local density of background sources. In each panel, the GCs are sorted in descending order from greatest to least excess. $1\sigma$ uncertainties inferred from bootstrapping are also plotted. Excess values of 0, without apparent uncertainties, are GCs with very small core or half-light radii.}\label{fig:rcrh_ex}
\end{figure*}

These core and half-light excesses are shown in Figure \ref{fig:rcrh_ex} and listed in Table \ref{tab:ex_vals}. The points represent median values of the boostrapped samples and the error bars standard $1\sigma$ uncertainties.

The behaviors differ somewhat between the
core and half-light samples. About 1/3 of the GCs show some evidence for an excess within a core radius, including four clusters that show excesses of at least 2 sources. One of these is M22, the first cluster in which our collaboration found candidate quiescent stellar-mass black holes \citep{straderetal2012}. Three of these four GCs also have relatively large core radii. About half the clusters show no apparent excess (see Figure \ref{fig:rcrh_ex}), in many cases due to their very small core radii. The combined overall significance of the core radius excess in this conservative bootstrapped analysis is about $4\sigma$.

The half-light radius excesses generally track those of the core radius, though in some cases with larger uncertainties or more extreme values. The two GCs with the largest half-light excesses are M4 and NGC 6544. The values for both of these GCs are boosted by the sensitivity correction, since there are a significant number of sources just above $5\sigma$ within the half-light radius. Each cluster would have a noticeable (but less extreme) excess were no correction applied.

\subsection{Interpreting the Radio Counts}

It is worth emphasizing that some radio sources have been individually established as cluster members, even in clusters where there is not a statistically significant excess of sources. For example, M10 contains a radio continuum source (M10-VLA1) which is slightly too faint to appear in a $5\sigma$ catalog for this cluster, but which optical spectroscopy of the counterpart proves is a cluster member \citep{shishkovskyetal2018}. Other clusters in the sample host radio pulsars, which are certainly radio continuum sources, but mostly too faint to be detectable at 5.0 GHz. The two clusters with the largest core excesses are M4 and M22, two of the nearest objects in our sample. It is clear that our survey is sensitivity limited and that even the VLA can only reveal the brighter part of the luminosity distribution of radio continuum sources in GCs. Hence our excess measurements represent conservative lower limits to the radio source population. A complete analysis of the luminosity function of radio sources in the MAVERIC sample, including both VLA and ATCA data, will occur in a future paper.

Since one of the central goals of MAVERIC is to constrain the population of stellar-mass black holes in GCs, we note that twelve of our sample GCs also appear in the study of \citet{weatherfordetal2019}, who combine models of how black hole populations affect cluster structure and mass segregation with observed data for a sample of Galactic GCs, theoretically predicting the number of black holes in each cluster. Of the twelve overlapping GCs, the two with the largest core excesses of radio sources, M22 and M55, are both in the top quartile of predicted black hole numbers. On the other hand, we see only mild evidence for a core radio source excess in M13, which has the largest predicted black hole population in the \citet{weatherfordetal2019} sample. Since most black holes are unlikely to be in mass-transferring binaries and hence observable as radio continuum sources, these comparisons are not straightforward to interpret. It would be valuable to obtain new, deep radio data for some of the GCs not in our current VLA sample that are suggested to host large black hole populations.

\subsection{Spectral Index Distribution}

An independent comparison of GC vs.~background sources is possible by considering the radio continuum spectral indices of the sources. Background AGN can have a range of spectral indices, depending on their orientation, size, and accretion state; those with large-scale jets typically show steeper $\alpha < -0.5$ spectra, while emission from the compact core of the AGN has a flatter spectrum ($\alpha \sim 0$) due to partially self-absorbed synchrotron radiation. Star forming galaxies typically have steeper spectra around $\alpha \sim -0.8$ arising from optically thin synchrotron radiation associated with supernovae
\citep{condon1992,dezottietal2010,huynhetal2015,tisanicetal2019}. The median spectral index observed in different surveys varies depending on the depth, selection method, and precise frequencies used, but is typically around $\alpha \sim$ --0.6 to --0.7, with a standard deviation of $\sim 0.4$--0.5 (e.g., \citealt{huynhetal2015,smolcicetal2017}).

\begin{figure}[!t]

\gridline{\fig{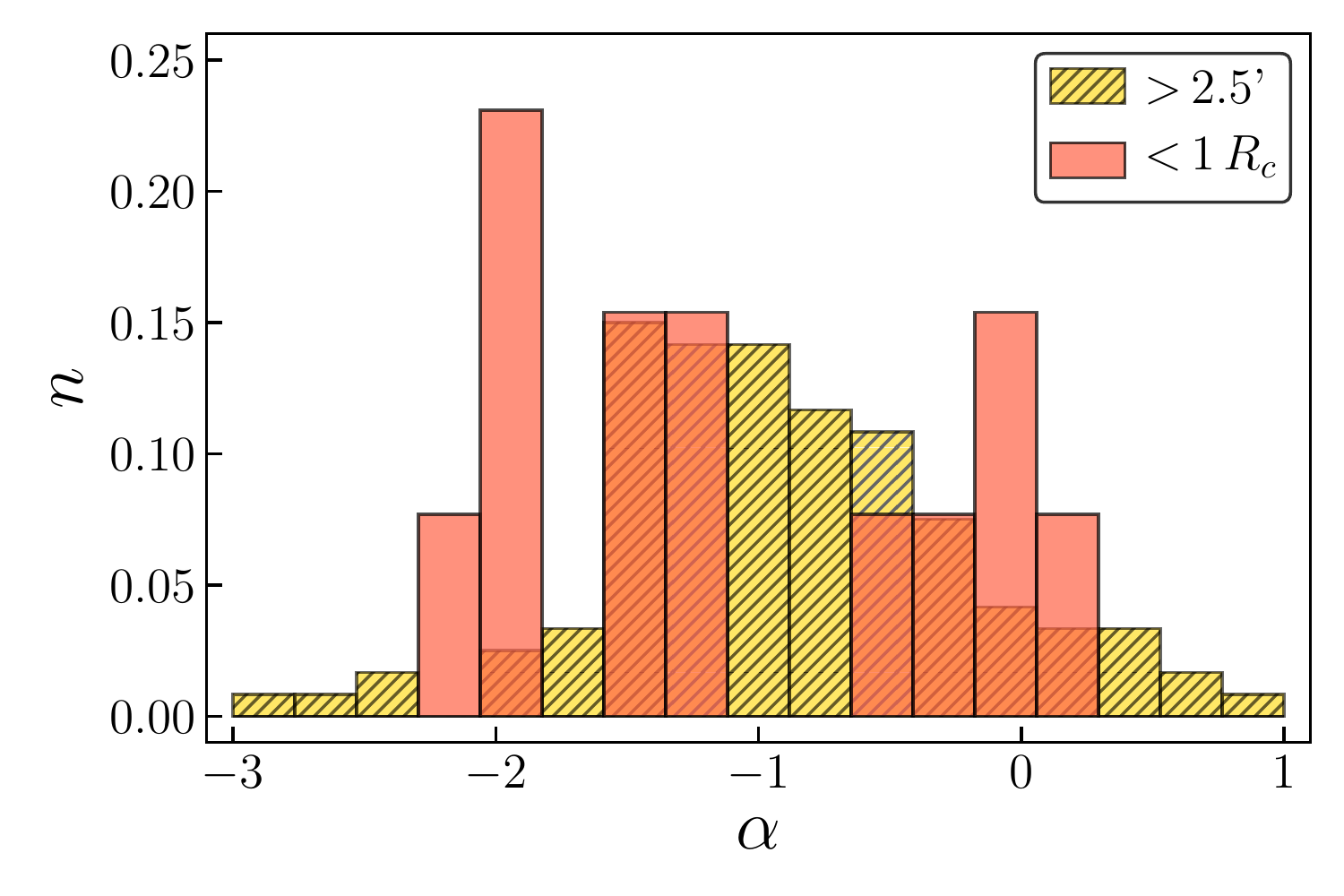}{0.45\textwidth}{\vspace{-0.45cm}\hspace{0.9cm}{\small{(a)}}}}

\vspace{-0.3cm}

\gridline{\fig{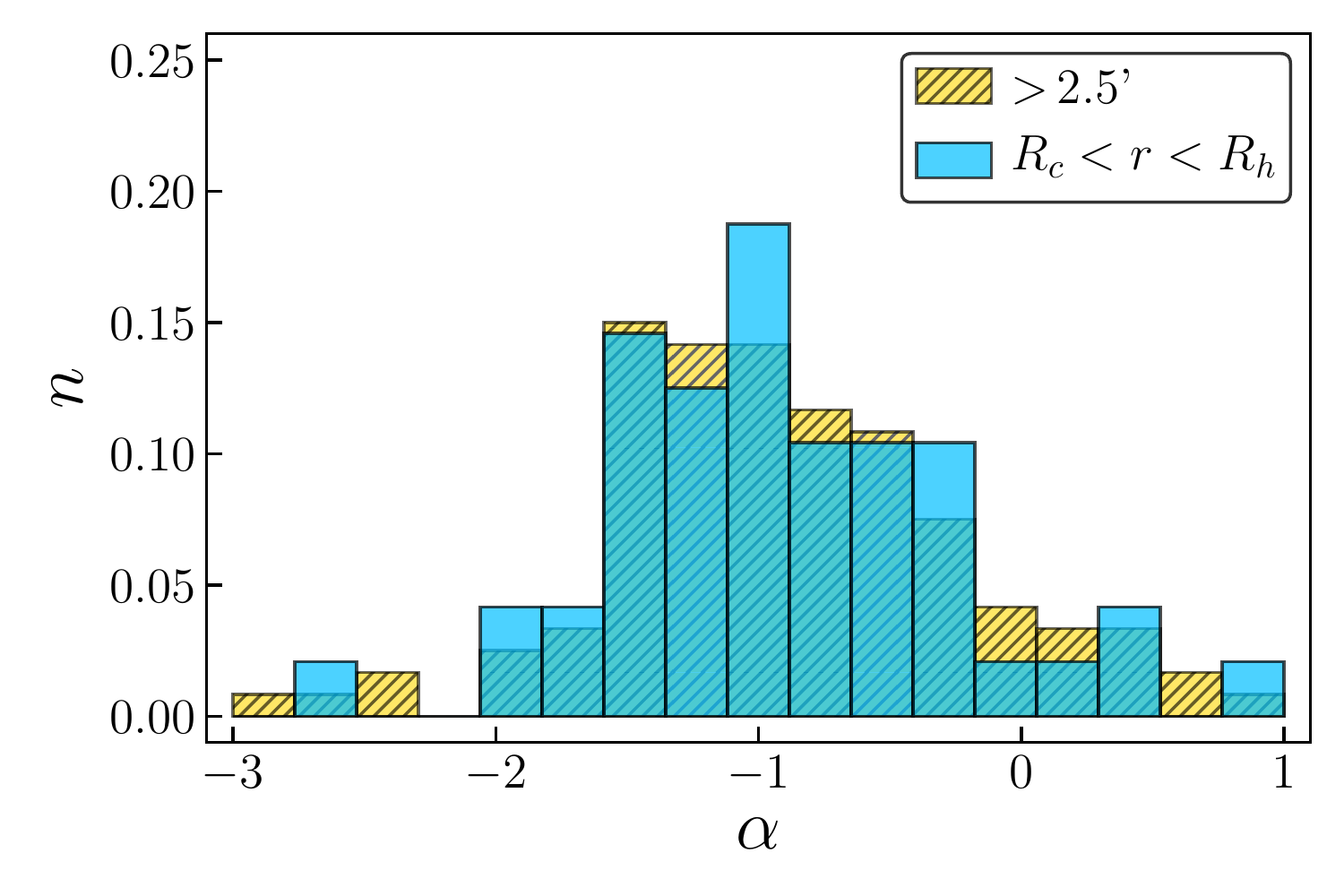}{0.45\textwidth}{\vspace{-0.45cm}\hspace{0.9cm}{\small{(b)}}}}

\vspace{-0.3cm}
\caption{Normalized distribution of spectral indices for sources inside the core (a: orange)  or between the half-light radius and the core (b: blue) vs. sources outside of 2.5{\arcmin}, our estimated background (yellow).}\label{fig:si_dist}

\end{figure}

In Figure \ref{fig:si_dist} we compare the spectral index distributions of sources within 1 core radius and outside 2.5$\arcmin$, representing ``clean" GC and background populations, respectively. Only those sources with spectral index uncertainties $< 0.5$ are plotted, and the distributions are normalized for comparison.

We find that the 120 background sources that meet these criteria have a broad distribution, as expected, with a median around $\alpha =$--0.9. By contrast, the 13 sources within the core radius have a bimodal distribution, with a very steep population with $\alpha$ mostly in the range --1 to --2, and a narrower peak centered around $\alpha \sim 0$ where there are few background sources. We tentatively identify the first population as millisecond pulsars, which show steep spectra (e.g., \citealt{batesetal2013}; \citealt{zhao2020}). The second, narrower peak is consistent with a population of binaries that show flat spectra, such as quiescent stellar-mass black holes, transitional millisecond pulsars, or even active binaries (see discussion in \citealt{shishkovskyetal2018,bahramianetal2018,millerjonesetal2015,chomiuketal2013,straderetal2012}). For example, one of these sources is M62-VLA1, which has been established as a low-mass X-ray binary in M62 \citep{chomiuketal2013}, though the nature of the accretor has yet to be definitely determined. 

The formal statistical support for a difference between the core and background populations is not well-established with these samples: an Anderson-Darling test finds $p=0.11$, which is not surprising given that the the number of sources within the core radii with well-measured spectral indices is relatively small. Hence the visually suggestive difference between the distribution of spectral indices for GC and background sources needs to be bolstered by additional future work.

The spectral index distribution of the 61 sources within the half-light radius (but outside the core) is more similar to the distribution of background sources (Figure \ref{fig:si_dist}b), suggesting that many (or even most) of the sources outside the core but within the half-light radius are indeed background sources. A similar Anderson-Darling test between the half-light radius and background samples gives $p=0.68$. Follow-up observations will be necessary to identify and classify the subset of these sources associated with the cluster rather than the background.

\section{Conclusions}

Here we have presented the initial results of our VLA radio continuum survey of Milky Way GCs. We observed 25 relatively nearby and massive clusters for $\sim$ 10 hrs each with the VLA at typical frequencies of 5.0 GHz and 7.2 GHz. These data represent the first deep high-resolution radio continuum survey of GCs as well as a sensitive C band $\sim 0.4$ deg$^2$ extragalactic survey. This paper presents radio source catalogs for each GC; in most cases these catalogs will represent a mixture of cluster and background sources. 

Through both source counts and spectral indices we found strong evidence for a population of radio sources associated with GCs. It is likely that a subset of these sources are steep-spectrum millisecond pulsars, while others represent compact or other kinds of binaries whose classification is not yet clear.

{Within the core radii, a conservative bootstrapped estimate suggests that just below half (23/54) of the 5.0 GHz sources are likely associated with the GCs. The corresponding estimate for the half-light radii is about 30\% (61/201), and the fraction is likely lower at larger radii.}

These catalogs represent an important but incomplete step in our efforts to understand the radio continuum source populations in GCs. Future papers from our group will use the full range of multi-wavelength information, including X-ray observations and optical photometry and spectroscopy, to determine and refine classifications of individual sources.

{New VLA data have recently been obtained for 7 additional GCs not discussed here, and an analysis of these data, as well as the ATCA part of the MAVERIC sample, will allow more secure conclusions about the properties and classification of cluster radio continuum sources.
In addition, despite our deep VLA images with $\sim 2$--3$\, \mu$Jy noise levels, we are still strongly limited by sensitivity: half of the VLA sources within the core radii have 5.0 GHz flux densities $< 15 \, \mu$Jy. Hence, surveys at these frequencies with future telescopes such as the Next Generation VLA \citep{Murphy18} should allow a severalfold increase in the number of detected cluster radio continuum sources. Finally, MAVERIC was optimized for the detection of flat-spectrum sources; there is substantial radio continuum discovery space at lower frequencies for steep spectrum GC sources such as pulsars with existing telescopes like the VLA and MeerKAT \citep{Booth09}.}

\acknowledgments
We thank the anonymous referee for a useful report that helped to improve the paper. We acknowledge support from NSF grants AST-1308124 and AST-1514763, NASA grant Chandra-GO8-19122X, and the Packard Foundation. The National Radio Astronomy Observatory is a facility of the National Science Foundation operated under cooperative agreement by Associated Universities, Inc. JCAM-J is the recipient of an Australian Research Council Future Fellowship (FT140101082), funded by the Australian government. COH is funded by NSERC Discovery Grant RGPIN-2016-04602. We acknowledge computational support from HPCC (High Performance Computing), Michigan State University.\\

\facilities{VLA, ATCA}
\software{AEGEAN \citep{2018hancock_aeg}, AIPS \citep{1985wells}, APLpy \citep{2012robit}, Astropy \citep{2013astropy}, CASA \citep{2007mcmullin}, Matplotlib \citep{hunter2007}, NumPy \citep{vanderwalt2011}, Pandas \citep{mckinney2010}, SAOImage DS9 \citep{joye2003}, SciPy \citep{virtanen2020}}.

\vspace{3.0cm}
\bibliography{src_cts_refs}

\clearpage
\newpage
\startlongtable
\begin{deluxetable}{llcc}
\tablecaption{Epochs of VLA Data \label{tab:vla_epoch}}
\tablehead{
\colhead{ID} & 
\colhead{Date} & 
\colhead{Obs. Time} & 
\colhead{Project Code} \\
\colhead{} & 
\colhead{} & 
\colhead{(hr)} & 
\colhead{} \\
}
\startdata
M2 & 2015 Jun 28 & 5 & 15A-100 \\
   & 2015 Jul 3 & 5 & 15A-100 \\
M3 & 2015 Jun 30 & 1 & 15A-100 \\ 
 & 2015 Jul 26 & 1 & 15A-100 \\
 & 2015 Aug 1 & 2 & 15A-100 \\
 & 2015 Aug 4 & 1 & 15A-100 \\
 & 2015 Aug 16 & 1 & 15A-100 \\
 & 2015 Aug 28 & 1 & 15A-100 \\ 
 & 2015 Aug 30 & 1 & 15A-100 \\
 & 2015 Sep 2 & 1 &  15A-100 \\
 & 2015 Sep 4 & 1 & 15A-100 \\
M4 & 2014 Feb 11 & 1 & 13B-014 \\
 & 2014 Feb 12 & 2 & 13B-014 \\
 & 2014 Feb 15 & 2 & 13B-014 \\
 & 2015 May 20 & 1 & 15A-100 \\
 & 2015 May 25 & 1 & 15A-100 \\
 & 2015 Jun 1 & 1 & 15A-100 \\
 & 2015 Jun 7 & 2 & 15A-100 \\
M5 & 2015 Jun 25 & 4 & 15A-100 \\
 & 2015 Jun 26 & 4 & 15A-100 \\
M9 & 2015 Jul 16 & 5 & 15A-100 \\
 & 2015 Jul 20 & 5 & 15A-100 \\
M10 & 2014 Feb 20 & 2 & 13B-014 \\
 & 2014 Mar 28 & 2 & 13B-014 \\
 & 2014 Apr 7 & 2 & 13B-014 \\
 & 2014 Apr 10 & 2 & 13B-014 \\
 & 2014 Apr 29 & 2 & 13B-014 \\
M12 & 2015 Jun 27 & 4 & 15A-100 \\
 & 2015 Jul 3 & 4 & 15A-100 \\
M13 & 2014 Feb 25 & 1 & 13B-014 \\
 & 2014 Mar 1 & 1 & 13B-014 \\
 & 2014 Mar 7 & 1 & 13B-014 \\
 & 2014 Apr 12 & 1 & 13B-014 \\
 & 2014 May 1 & 1 & 13B-014 \\
 & 2014 May 20 & 1 & 13B-014 \\
 & 2015 Jun 18 & 2.5 & 15A-100 \\
 & 2015 Jul 1 & 2.5 & 15A-100 \\
M14 & 2015 Jul 10 & 5 & 15A-100 \\
 & 2015 Jul 12 & 5 & 15A-100 \\
M19 & 2011 May 19 & 5 & 10C-109 \\
 	& 2011 May 20 & 2.5 & 10C-109 \\  
 	& 2011 May 21 & 2.5 & 10C-109 \\ 
M22 & 2011 May 22 & 2.5 & 10C-109 \\
 	& 2011 May 23 & 2.5 & 10C-109 \\  
 	& 2011 May 25 & 2.5 & 10C-109 \\
 	& 2011 May 29 & 2.5 & 10C-109 \\
M28 & 2014 May 2 & 1 & 13B-014 \\
 	& 2014 May 4 & 2 & 13B-014 \\
 	& 2014 May 19 & 2 & 13B-014 \\
 	& 2014 May 23 & 1 & 13B-014 \\
 	& 2014 May 29 & 1 & 13B-014 \\
 	& 2014 May 30 & 2 & 13B-014 \\
M30 & 2015 Jun 23 & 3.3 & 15A-100 \\
 	& 2015 Jul 4 & 3.3 &  15A-100 \\
 	& 2015 Jul 5 & 3.3 &  15A-100 \\
M54 & 2015 Jun 9 & 2 & 15A-100 \\
 	& 2015 Jun 10 & 1 & 15A-100 \\  
 	& 2015 Jun 12 & 2 & 15A-100 \\  
 	& 2015 Jun 17 & 1 & 15A-100 \\ 
M55 & 2015 May 20 & 1 & 15A-100 \\
 	& 2015 May 22 & 1 & 15A-100 \\
 	& 2015 Jun 1 & 1 & 15A-100 \\
 	& 2015 Jun 2 & 1 & 15A-100 \\
 	& 2015 Jun 4 & 1 & 15A-100 \\
 	& 2015 Jun 5 & 1 & 15A-100 \\
 	& 2015 Jun 6 & 2 & 15A-100 \\
M62 & 2012 Sep 10 & 1 & 12B-073 \\
 	& 2012 Sep 11 & 3.75 & 12B-073 \\
 	& 2012 Sep 14 & 1.75 & 12B-073 \\
 	& 2012 Sep 15 & 1.75 & 12B-073 \\
 	& 2012 Sep 16 & 1.75 & 12B-073 \\
M92 & 2015 Jun 19 & 3.3 & 15A-100 \\
	 & 2015 Jun 24 & 3.3 & 15A-100 \\
	 & 2015 Jul 4 & 3.3 & 15A-100 \\
M107 & 2015 Jul 8 & 5 & 15A-100 \\
	 & 2015 Jul 9 & 5 & 15A-100 \\
NGC 6304 & 2015 Jun 9 & 3 & 15A-100 \\
 & 2015 Jun 12 & 1 & 15A-100 \\
 & 2015 Jun 17 & 1 & 15A-100 \\
NGC 6325 & 2015 Jul 23 & 2 & 15A-100 \\
 & 2015 Jul 29 & 4 & 15A-100 \\
 & 2015 Aug 1 & 2 & 15A-100 \\
 & 2015 Aug 3 & 4 & 15A-100 \\
NGC 6440 & 2014 May 5 & 1 & 13B-014 \\
 & 2014 May 12 & 1 & 13B-014 \\
 & 2014 May 16 & 1 & 13B-014 \\
 & 2014 Jun 1 & 2 & 13B-014 \\
 & 2015 Jul 6 & 5 & 15A-100 \\
NGC 6539  & 2015 Jul 2 & 5 & 15A-100 \\
 & 2015 Jul 4 & 5 & 15A-100 \\
NGC 6544 & 2014 May 3 & 1 & 13B-014 \\
 & 2014 May 5 & 1 & 13B-014 \\
 & 2014 May 6 & 1 & 13B-014 \\
 & 2014 May 10 & 1 & 13B-014 \\
 & 2014 May 13 & 1 & 13B-014 \\
 & 2014 May 22 & 1 & 13B-014 \\
 & 2014 May 31 & 2 & 13B-014 \\
 & 2014 Jun 2 & 1 & 13B-014 \\
 & 2015 Jul 11 & 1 & 15A-100 \\
NGC 6712 & 2014 Apr 5 & 2 & 13B-014 \\
 & 2014 May 5 & 1 & 13B-014 \\
 & 2014 May 8 & 2 & 13B-014 \\
 & 2014 May 9 & 1 & 13B-014 \\
 & 2014 May 13 & 2 & 13B-014 \\
 & 2014 May 18 & 1 & 13B-014 \\
 & 2014 May 20 & 1 & 13B-014 \\
NGC 6760 & 2015 Jul 1 & 5 & 15A-100 \\
 & 2015 Jul 7 & 5 & 15A-100 \\
\enddata
\end{deluxetable}

\newpage
\begin{deluxetable}{lrrrrrrcccccccc}
\tabletypesize{\small}
\tablewidth{0pt} 
\rotate
\tablecaption{Information for our cluster sample and images\label{tab:vla_im}}
 \tablehead{
 \colhead{ID}           & 
 \colhead{R.A. (J2000)}          &
 \colhead{Dec. (J2000)}          & 
 \colhead{Distance} &
 \colhead{Core} &
 \colhead{Half-light} &
\colhead{time} &
\colhead{on src} &
\colhead{$\nu_{\rm low}$} &
\colhead{beam$_{\rm low}$} &
\colhead{rms$_{\rm low}$} &
\colhead{$\nu_{\rm high}$} &
\colhead{beam$_{\rm high}$} &
\colhead{rms$_{\rm high}$}  \\
 \colhead{} &
 \colhead{(h:m:s)} &
 \colhead{(\phn{\arcdeg}:\phn{\arcmin}:\phn{\arcsec})} &
 \colhead{(kpc)} &
\colhead{radius (\phn{\arcsec})}&
\colhead{radius (\phn{\arcsec})}&
 \colhead{(hr)} &
 \colhead{(hr)} &
 \colhead{(GHz)} &
 \colhead{} &
 \colhead{$\mu$Jy beam$^{-1}$} &
 \colhead{(GHz)} &
 \colhead{} &
 \colhead{$\mu$Jy beam$^{-1}$} \\
 }
\startdata
M2 & 21:33:26.96 & --00:49:22.90 & 11.5 & 19.2 & 63.6 & 10 & 8.3 & 5.2 & $0.46\arcsec \times 0.38\arcsec$ & 1.8 & 7.2 & $0.31\arcsec \times 0.26\arcsec$ & 1.7 \\
M3 & 13:42:11.38 & +28:22:39.10 & 10.1 & 22.2 & 138.6 & 10 & 7.1 & 5.0 & $0.58\arcsec \times 0.46\arcsec$ & 2.8 & 7.0 & $0.37\arcsec \times 0.29\arcsec$ & 2.1 \\
M4 & 16:23:35.03 & --26:31:33.80 & 1.8 & 69.6 & 259.8 & 10 & 6.9 & 5.0 & $1.18\arcsec \times 0.88\arcsec$ & 2.3 & 7.2 & $0.84\arcsec \times 0.63\arcsec$ & 2.1 \\
M5 & 15:18:33.21 & +02:04:51.80 & 7.7 & 26.4 & 106.2 & 8 & 6.6 & 5.2 & $0.47\arcsec \times 0.39\arcsec$ & 1.8 & 7.2 & $0.33\arcsec \times 0.28\arcsec$ & 1.8 \\
M9 & 17:19:11.78 & --18:30:58.50 & 7.8 & 27.0 & 57.6 & 10 & 8.2 & 5.0 & $0.63\arcsec \times 0.41\arcsec$ & 1.7 & 7.0 & $0.44\arcsec \times 0.31\arcsec$ & 1.7 \\
M10 & 16:57:08.92 & --04:05:58.00 & 4.4 & 46.2 & 117.0 & 10 & 7.9 & 5.0 & $0.75\arcsec \times 0.36\arcsec$ & 2.5 & 7.4 & $0.53\arcsec \times 0.26\arcsec$ & 1.9 \\
M12 & 16:47:14.18 & --01:56:54.70 & 5.2 & 47.4 & 106.2 & 8 & 6.6 & 5.2 & $0.49\arcsec \times 0.40\arcsec$ & 2.0 & 7.2 & $0.39\arcsec \times 0.27\arcsec$ & 1.8 \\
M13 & 16:41:41.21 & +36:27:35.60 & 7.6 & 37.2 & 101.4 & 11 & 7.6 & 5.0 & $0.54\arcsec \times 0.45\arcsec$ & 2.0 & 7.2 & $0.35\arcsec \times 0.28\arcsec$ & 2.1 \\
M14 & 17:37:36.10 & --03:14:45.30 & 9.3 & 47.4 & 78.0 & 10 & 8.4 & 5.3 & $0.46\arcsec \times 0.40\arcsec$ & 1.8 & 7.2 & $0.35\arcsec \times 0.29\arcsec$ & 1.7 \\
M19 & 17:02:37.80 & --26:16:04.70 & 8.2 & 25.8 & 79.2 & 10 & 7.4 & 5.0 & $1.28\arcsec \times 0.96\arcsec$ & 2.3 & 6.8 & $0.95\arcsec \times 0.72\arcsec$ & 2.0 \\
M22 & 18:36:23.94 & --23:54:17.10 & 3.1 & 79.8 & 201.6 & 10 & 7.3 & 5.0 & $1.54\arcsec \times 0.81\arcsec$ & 2.4 & 6.8 & $1.14\arcsec \times 0.59\arcsec$ & 2.0 \\
M28 & 18:24:32.73 & --24:52:13.00 & 5.5 & 14.4 & 118.2 & 9 & 6.0 & 5.0 & $0.94\arcsec \times 0.40\arcsec$ & 2.5 & 7.4 & $0.66\arcsec \times 0.27\arcsec$ & 2.1 \\
M30 & 21:40:22.12 & --23:10:47.50 & 8.6 & 3.6 & 61.8 & 10 & 8.0 & 4.9 & $0.74\arcsec \times 0.39\arcsec$ & 1.7 & 7.0 & $0.52\arcsec \times 0.28\arcsec$ & 1.6 \\
M54 & 18:55:03.33 & --30:28:47.50 & 23.9 & 5.4 & 49.2 & 6 & 3.8 & 5.1 & $1.09\arcsec \times 0.44\arcsec$ & 3.4 & 7.1 & $0.74\arcsec \times 0.30\arcsec$ & 2.8 \\
M55 & 19:39:59.71 & --30:57:53.10 & 5.7 & 108.0 & 169.8 & 8 & 5.1 & 5.1 & $1.15\arcsec \times 0.75\arcsec$ & 2.3 & 7.1 & $0.83\arcsec \times 0.62\arcsec$ & 2.2 \\
M62 & 17:01:12.98 & --30:06:49.00 & 6.7 & 13.2 & 55.2 & 10 & 7.0 & 5.0 & $1.53\arcsec \times 1.16\arcsec$ & 3.2 & 7.4 & $0.95\arcsec \times 0.71\arcsec$ & 2.2 \\
M92 & 17:17:07.43 & +43:08:09.30 & 8.9 & 15.6 & 61.2 & 10 & 7.9 & 5.0 & $0.35\arcsec \times 0.38\arcsec$ & 1.7 & 7.1 & $0.33\arcsec \times 0.27\arcsec$ & 1.6 \\
M107 & 16:32:31.86 & --13:03:13.60 & 6.1 & 33.6 & 103.8 & 10 & 8.2 & 5.0 & $0.63\arcsec \times 0.42\arcsec$ & 2.3 & 7.2 & $0.41\arcsec \times 0.29\arcsec$ & 2.2 \\
N6304 & 17:14:32.25 & --29:27:43.30 & 5.9 & 12.6 & 85.2 & 5 & 3.2 & 5.0 & $0.91\arcsec \times 0.56\arcsec$ & 3.8 & 7.1 & $0.65\arcsec \times 0.44\arcsec$ & 3.9 \\
N6325 & 17:17:59.21 & --23:45:57.60 & 6.5 & 1.8 & 37.8 & 12 & 9.0 & 5.0 & $0.68\arcsec \times 0.41\arcsec$ & 1.9 & 7.1 & $0.52\arcsec \times 0.31\arcsec$ & 2.0 \\
N6440 & 17:48:52.70 & --20:21:36.90 & 8.5 & 8.4 & 28.8 & 10 & 7.7 & 5.0 & $0.90\arcsec \times 0.60\arcsec$ & 2.7 & 7.1 & $0.65\arcsec \times 0.40\arcsec$ & 2.6 \\
N6539 & 18:04:49.68 & --07:35:09.10 & 7.8 & 22.8 & 102.0 & 10 & 8.2 & 5.2 & $0.51\arcsec \times 0.39\arcsec$ & 1.7 & 7.2 & $0.39\arcsec \times 0.28\arcsec$ & 2.2 \\
N6544 & 18:07:20.12 & --24:59:53.60 & 3.0 & 3.0 & 72.6 & 10 & 6.2 & 5.0 & $1.02\arcsec \times 0.39\arcsec$ & 2.5 & 7.4 & $0.69\arcsec \times 0.26\arcsec$ & 2.1 \\
N6712 & 18:53:04.30 & --08:42:22.00 & 8.0 & 45.6 & 79.8 & 10 & 7.2 & 5.0 & $0.66\arcsec \times 0.39\arcsec$ & 2.3 & 7.4 & $0.46\arcsec \times 0.26\arcsec$ & 2.1 \\
N6760 & 19:11:12.01 & +01:01:49.70 & 7.4 & 20.4 & 76.2 & 10 & 8.2 & 5.1 & $0.52\arcsec \times 0.42\arcsec$ & 2.2 & 7.2 & $0.37\arcsec \times 0.29\arcsec$ & 2.0 \\
\enddata
\tablecomments{The sources for the photometric centers, distances, and structural parameters are as in \citet{tremouetal2018}.}
\end{deluxetable}

\clearpage
\newpage
\begin{deluxetable}{lccccrrrrcccr}
\tabletypesize{\small}
\tablecolumns{13} 
\tablewidth{0pt} 
\rotate
\tablecaption{Radio Continuum Sources \label{tab:vla_src}}
\tablehead{
\colhead{ID} & 
\colhead{R.A.\tablenotemark{a}} & 
\colhead{Dec.\tablenotemark{a}} & 
\colhead{R.A. unc.} & 
\colhead{Dec. unc.} & 
\colhead{$S_{5}$\tablenotemark{b}} &
\colhead{$S_{5}$ unc.} &
\colhead{$S_{7}$\tablenotemark{c}} & 
\colhead{$S_{7}$ unc.} & 
\colhead{note\tablenotemark{d}} & 
\colhead{radius\tablenotemark{e}} & 
\colhead{loc.\tablenotemark{f}} & 
\colhead{$\alpha$\tablenotemark{g}} \\
\colhead{} & 
\colhead{(h:m:s)} &
\colhead{($^{\circ}$:\arcmin:\arcsec)} &
\colhead{(\arcsec)} &
\colhead{(\arcsec)} & 
\colhead{($\mu$Jy)} &
\colhead{($\mu$Jy)} & 
\colhead{($\mu$Jy)} &
\colhead{($\mu$Jy)} &
\colhead{}   &
\colhead{(\arcmin)}&
\colhead{}  &
\colhead{} \\ }
\startdata
M2-VLA1 & 21:33:36.358 & --00:51:10.14 & \nodata & \nodata & 19013.3 & 5.6 & 12724.0 & 4.8 & ext. & 2.95 & \nodata & $-1.23_{-0.01}^{+0.01}$ \\
M2-VLA2 & 21:33:23.668 & --00:46:46.08 & 0.03 & 0.03 & 190.7 & 2.2 & 156.4 & 2.7 & \nodata & 2.74 & \nodata & $-0.61_{-0.06}^{+0.06}$ \\
M2-VLA3 & 21:33:32.016 & --00:51:34.79 & 0.03 & 0.03 & 134.4 & 2.1 & 119.4 & 2.7 & \nodata & 2.54 & \nodata & $-0.36_{-0.09}^{+0.08}$ \\
M2-VLA4 & 21:33:29.770 & --00:47:43.41 & 0.03 & 0.03 & 97.6 & 2.0 & 104.1 & 2.1 & \nodata & 1.80 & \nodata & $+0.20_{-0.09}^{+0.09}$ \\
M2-VLA5 & 21:33:26.964 & --00:46:45.88 & 0.03 & 0.03 & 90.9 & 2.2 & 70.1 & 2.7 & \nodata & 2.62 & \nodata & $-0.80_{-0.14}^{+0.14}$ \\
M2-VLA6 & 21:33:17.915 & --00:51:56.40 & 0.03 & 0.03 & 86.2 & 2.7 & 141.0 & 3.9 & \nodata & 3.41 & \nodata & $+1.42_{-0.09}^{+0.06}$ \\
M2-VLA7 & 21:33:34.476 & --00:46:49.45 & 0.03 & 0.03 & 76.7 & 2.4 & 83.3 & 3.4 & \nodata & 3.17 & \nodata & $+0.25_{-0.16}^{+0.16}$ \\
M2-VLA8 & 21:33:23.387 & --00:46:43.19 & 0.03 & 0.03 & 52.7 & 2.2 & 40.8 & 2.8 & \nodata & 2.81 & \nodata & $-0.80_{-0.25}^{+0.24}$ \\
M2-VLA9 & 21:33:36.614 & --00:51:12.55 & 0.03 & 0.03 & 43.8 & 5.4 & 50.1 & 4.4 & \nodata & 3.03 & \nodata & $+0.43_{-0.46}^{+0.47}$ \\
M2-VLA10 & 21:33:28.879 & --00:49:43.34 & 0.03 & 0.03 & 41.5 & 1.9 & 30.4 & 1.8 & \nodata & 0.59 & $r_{h}$ & $-0.95_{-0.24}^{+0.23}$ \\
M2-VLA11 & 21:33:21.993 & --00:52:15.52 & 0.03 & 0.03 & 34.3 & 2.5 & 27.9 & 3.5 & \nodata & 3.13 & \nodata & $-0.66_{-0.48}^{+0.44}$ \\
M2-VLA12 & 21:33:18.304 & --00:48:24.60 & 0.03 & 0.03 & 33.9 & 2.1 & 53.9 & 2.4 & \nodata & 2.37 & \nodata & $+1.31_{-0.18}^{+0.13}$ \\
M2-VLA13 & 21:33:24.849 & --00:46:30.12 & 0.03 & 0.03 & 27.1 & 2.3 & 25.8 & 3.0 & \nodata & 2.93 & \nodata & $-0.17_{-0.46}^{+0.44}$ \\
M2-VLA14 & 21:33:21.935 & --00:49:45.76 & 0.03 & 0.03 & 27.0 & 1.9 & 25.0 & 1.9 & \nodata & 1.31 & \nodata & $-0.24_{-0.32}^{+0.32}$ \\
M2-VLA15 & 21:33:36.424 & --00:51:07.87 & 0.03 & 0.03 & 24.4 & 5.7 & 33.2 & 4.6 & art.? & 2.94 & \nodata & $+0.70_{-0.67}^{+0.53}$ \\
M2-VLA16 & 21:33:36.316 & --00:51:00.79 & 0.04 & 0.05 & 23.3 & 3.7 & <11.7 & -1.2 & art.? & 2.85 & \nodata & $<0.5$ \\
M2-VLA17 & 21:33:36.680 & --00:49:11.56 & 0.03 & 0.03 & 22.0 & 2.4 & 15.3 & 2.6 & \nodata & 2.44 & \nodata & $-1.18_{-0.68}^{+0.62}$ \\
M2-VLA18 & 21:33:24.414 & --00:49:43.52 & 0.03 & 0.03 & 20.2 & 1.9 & 11.8 & 1.8 & \nodata & 0.72 & $r_{h}$ & $-1.71_{-0.59}^{+0.55}$ \\
M2-VLA19 & 21:33:32.957 & --00:48:16.40 & 0.03 & 0.03 & 18.2 & 2.0 & 10.9 & 2.0 & \nodata & 1.86 & \nodata & $-1.64_{-0.71}^{+0.66}$ \\
M2-VLA20 & 21:33:36.371 & --00:50:29.77 & 0.04 & 0.05 & 17.9 & 2.8 & <8.4 & 0.3 & \nodata & 2.60 & \nodata & $<0.2$ \\
M2-VLA21 & 21:33:13.408 & --00:50:07.37 & 0.04 & 0.05 & 17.1 & 2.7 & <12.0 & -0.1 & \nodata & 3.47 & \nodata & $<1.1$ \\
M2-VLA22 & 21:33:16.108 & --00:48:43.84 & 0.03 & 0.03 & 16.4 & 2.2 & 11.5 & 2.9 & \nodata & 2.79 & \nodata & $-1.23_{-1.00}^{+0.88}$ \\
M2-VLA23 & 21:33:37.458 & --00:48:53.07 & 0.03 & 0.03 & 16.3 & 2.4 & 15.6 & 2.7 & \nodata & 2.67 & \nodata & $-0.19_{-0.75}^{+0.70}$ \\
M2-VLA24 & 21:33:14.770 & --00:50:10.13 & 0.03 & 0.03 & 16.0 & 2.5 & 15.0 & 3.5 & \nodata & 3.15 & \nodata & $-0.32_{-0.99}^{+0.84}$ \\
M2-VLA25 & 21:33:34.411 & --00:48:52.10 & 0.03 & 0.03 & 15.1 & 2.1 & 9.4 & 2.1 & \nodata & 1.93 & \nodata & $-1.54_{-0.87}^{+0.81}$ \\
... & & & & & & & & & & & & \\
\enddata
\tablenotetext{a}{ICRS position at epoch of observation.}
\tablenotetext{b}{Flux density in the lower subband.}
\tablenotetext{c}{Flux density in the higher subband.}
\tablenotetext{d}{Whether the source is extended or a possible imaging artifact.}
\tablenotetext{e}{Projected radius from the cluster center.}
\tablenotetext{f}{Whether the source is within the core ($r_c$) or half-light ($r_h$) radius.}
\tablenotetext{g}{Spectral index $\alpha$ of source, for $S \propto \nu^\alpha$}
\end{deluxetable}

\clearpage
\newpage
\startlongtable
\begin{deluxetable}{lllcccc}
\tabletypesize{\small}
\tablecaption{Source Counts at 5.0 and 7.2 GHz. \label{tab:scounts}}
\tablehead{
\colhead{$S_{\nu}$\tablenotemark{a}}   &
\colhead{lim. (low)\tablenotemark{b}}  &
\colhead{lim. (high)\tablenotemark{c}}  &
\colhead{$N_{5.0}$\tablenotemark{d}}   &
\colhead{$N_{7.2}$\tablenotemark{e}}   &
\colhead{${S_{\nu}^{2.5}} \, dN_{5.0}/dS$\tablenotemark{f}}  &
\colhead{${S_{\nu}^{2.5}} \, dN_{7.2}/dS$\tablenotemark{g}} \\
\colhead{($\mu$Jy)}   &
\colhead{($\mu$Jy)}   &
\colhead{($\mu$Jy)}   &
\colhead{}   &
\colhead{}   &
\colhead{(Jy$^{1.5}$ sr$^{-1}$)}   &
\colhead{(Jy$^{1.5}$ sr$^{-1}$)} \
}
\startdata
9.15 & 7.90 & 10.61 & 35 & 32 &  $0.41_{-0.07}^{+0.08}$  & $0.52^{+0.11}_{-0.09}$ \\
12.3 & 10.61 & 14.25 & 162 & 127 & $0.67^{+0.06}_{-0.05}$ & $0.80^{+0.08}_{-0.07}$ \\
16.5 & 14.25 & 19.14 & 211 & 159 & $0.66^{+0.05}_{-0.05}$ & $0.74^{+0.06}_{-0.06}$ \\
22.2 & 19.14 & 25.72 & 157 & 149 & $0.62^{+0.05}_{-0.05}$ & $0.74^{+0.07}_{-0.06}$ \\
29.8 & 25.72 & 34.55 & 96 & 94 & $0.58^{+0.07}_{-0.06}$ & $0.60^{+0.07}_{-0.06}$ \\
40 & 34.55 & 46.42 & 67 & 56 & $0.63^{+0.09}_{-0.08}$ & $0.53^{+0.08}_{-0.07}$ \\
53.8 & 46.42 & 62.36 & 33 & 28 & $0.48^{+0.10}_{-0.08}$ & $0.41^{+0.09}_{-0.08}$ \\
72.3 & 62.36 & 83.77 & 33 & 18 & $0.75^{+0.16}_{-0.13}$ & $0.41^{+0.12}_{-0.10}$ \\
97.1 & 83.77 & 112.53 & 21 & 18 & $0.74^{+0.20}_{-0.16}$ & $0.64^{+0.19}_{-0.15}$ \\
130 & 112.53 & 151.18 & 19 & 13 & $1.04^{+0.30}_{-0.24}$ & $0.71^{+0.26}_{-0.20}$ \\
175 & 151.18 & 203.09 & 10 & 6 & $0.86^{+0.37}_{-0.27}$ & $0.51^{+0.31}_{-0.20}$ \\
235 & 203.09 & 272.83 & 6 & 8 & $0.80^{+0.48}_{-0.32}$ & $1.07^{+0.53}_{-0.37}$ \\
316 & 272.83 & 366.52 & 8 & 3 & $1.67^{+0.82}_{-0.58}$ & $0.62^{+0.61}_{-0.34}$ \\
425 & 366.52 & 492.39 & 4 & 3 & $1.30^{+1.02}_{-0.62}$ & $0.97^{+0.94}_{-0.53}$ \\
571 & 492.39 & 661.47 & 3 & 4 & $1.52^{+1.47}_{-0.82}$ & $2.03^{+1.60}_{-0.97}$ \\
767 & 661.47 & 888.62 & 2 & 0 & $1.58^{+2.07}_{-1.02}$ & --- \\
1030 & 888.62 & 1193.8 & 3 & 3 & $3.68^{+3.57}_{-2.00}$ & $3.68^{+3.57}_{-2.00}$ \\
1380 & 1193.8 & 1603.7 & 1 & 3 & $1.90^{+4.34}_{-1.57}$ & $5.69^{+5.52}_{-3.09}$ \\
1860 & 1603.7 & 2154.4 & 1 & 0 & $2.98^{+6.81}_{-2.47}$ & --- \\
2500 & 2154.4 & 2894.3 & 3 & 1 & $13.92^{+13.50}_{-7.55}$ & $4.64^{+10.62}_{-3.85}$ \\
3350 & 2894.3 & 3888.2 & 0 & 2 & --- & $14.37^{+18.87}_{-9.25}$ \\
4510 & 3888.2 & 5223.4 & 2 & 0 & $22.40^{+29.43}_{-14.43}$ & --- \\
6050 & 5223.4 & 7017.0 & 0 & 0 & --- & --- \\
8130 & 7017.0 & 9426.7 & 0 & 0 & --- & --- \\
10900 & 9426.7 & 12663.8 & 0 & 0 & --- & --- \\
14700 & 12663.8 & 17012.5 & 0 & 1 & --- & $66.21^{+151.44}_{-54.85}$ \\
19700 & 17012.5 & 22854.6 & 1 & 0 & $102.50^{+234.44}_{-84.92}$ & --- \\
\enddata
\tablenotetext{a}{Characteristic flux value of the bin (geometric mean).}
\tablenotetext{b}{Lower limit of bin.}
\tablenotetext{c}{Upper limit of bin.}
\tablenotetext{d}{Raw 5.0 GHz counts.}
\tablenotetext{e}{Raw 7.2 GHz counts.}
\tablenotetext{f}{Euclidian-normalized 5.0 GHz differential source counts.}
\tablenotetext{g}{Euclidian-normalized 7.2 GHz differential source counts.}
\end{deluxetable}

\clearpage
\newpage
\begin{deluxetable}{lrr}
\tabletypesize{\small}
\tablecaption{5.0 GHz Radio Source Excesses
\label{tab:ex_vals}}
\tablehead{
\colhead{ID} &
\colhead{Excess ($r<r_c$)} &
\colhead{Excess ($r<r_h$)} 
}
\startdata
M2 & $-0.27\pm0.07$ & $2.77\pm2.60$  \\
M3 & $-0.24\pm0.06$ & $2.73\pm4.26$  \\
M4 & $5.37\pm2.57$ & $29.47\pm10.16$  \\
M5 & $-0.53\pm0.12$ & $-4.29\pm3.04$  \\
M9 & $-0.09\pm1.02$ & $-2.69\pm1.87$  \\
M10 & $-1.34\pm0.36$ & $2.28\pm4.25$  \\
M12 & $-1.38\pm1.14$ & $-2.36\pm3.87$  \\
M13 & $1.31\pm1.38$ & $3.54\pm3.22$  \\
M14 & $1.48\pm2.03$ & $7.34\pm3.86$  \\
M19 & $0.64\pm0.97$ & $2.95\pm2.81$  \\
M22 & $6.05\pm3.65$ & $-0.97\pm6.92$  \\
M28 & $1.73\pm1.40$ & $-7.99\pm5.09$  \\
M30 & $-0.016\pm0.003$ & $-1.96\pm2.09$  \\
M54 & $-0.011\pm0.003$ & $-0.03\pm1.02$  \\
M55 & $4.50\pm3.96$ & $6.20\pm6.36$  \\
M62 & $0.89\pm0.95$ & $2.09\pm2.12$  \\
M92 & $-0.29\pm0.05$ & $-0.61\pm2.19$  \\
M107 & $0.68\pm1.47$ & $1.27\pm4.48$  \\
NGC 6304 & $-0.05\pm0.02$ & $-1.62\pm1.25$  \\
NGC 6325 & $-0.002\pm0.001$ & $-0.20\pm1.07$  \\
NGC 6440 & $0.94\pm1.02$ & $1.31\pm1.43$  \\
NGC 6539 & $3.78\pm1.96$ & $4.36\pm3.28$  \\
NGC 6544 & $-0.007\pm0.001$ & $20.14\pm10.22$  \\
NGC 6712 & $0.49\pm1.42$ & $-2.47\pm1.71$  \\
NGC 6760 & $-0.46\pm0.10$ & $-0.46\pm2.96$  \\
\enddata
\tablecomments{These are the sensitivity-corrected
``excesses" of 5.0 GHz radio sources within the core or half-light radius of the GC compared to the scaled local density of background sources.}
\end{deluxetable}

\end{document}